\documentclass[apjl]{emulateapj}
\usepackage{amsmath}
\usepackage{txfonts}
\usepackage{natbib}
\usepackage{graphicx}
\usepackage{color}

\newcommand{\msun}{{\ensuremath{M_\odot}}}
\newcommand{\msf}{{\ensuremath{M_\mathrm{4}}}}
\newcommand{\scc}{{\ensuremath{\mu_{4}}}} 

\newcommand{\xsi}{{\ensuremath{\xi_{1.5}}}}
\newcommand{\XSI}{{\ensuremath{\xi_{2.5}}}}

\shorttitle{Explosion criterion of massive stars}

\shortauthors{Ertl et al.}

\begin{document}

\title{A two-parameter criterion for classifying the explodability of massive
stars by the neutrino-driven mechanism}

\author{T.~Ertl\altaffilmark{1,2}, H.-Th.~Janka\altaffilmark{1},
S.~E.\ Woosley\altaffilmark{3}, T.~Sukhbold\altaffilmark{3},
and M.~Ugliano\altaffilmark{4}}

\altaffiltext{1}{Max-Planck-Institut f\"ur Astrophysik, Karl-Schwarzschild-Str.~1,
85748 Garching, Germany}
\altaffiltext{2}{Physik Department, Technische Universit\"at M\"unchen,
James-Franck-Stra\ss e 1, 85748 Garching, Germany}
\altaffiltext{3}{Department of Astronomy and Astrophysics, University of
California, Santa Cruz, CA 95064}
\altaffiltext{4}{Institut f\"ur Kernphysik, Technische Universit\"at Darmstadt,
Schlossgartenstr.~2, 64289 Darmstadt, Germany}
\email{tertl@mpa-garching.mpg.de, thj@mpa-garching.mpg.de}

\begin{abstract}
Thus far, judging the fate of a massive star (either a neutron star (NS)
or a black hole) solely by its structure prior to core collapse has been
ambiguous. Our work and previous attempts find a non-monotonic variation
of successful and failed supernovae with zero-age main-sequence mass, for
which no single structural parameter can serve as a good predictive measure.
However, we identify two parameters computed from the pre-collapse structure
of the progenitor, which in combination allow for a clear separation of
exploding and non-exploding cases with only few exceptions ($\sim$1--2.5\%)
in our set of 621 investigated stellar models. One parameter is $M_4$,
defining the normalized enclosed mass for a dimensionless entropy per 
nucleon of $s=4$,
and the other is 
$\mu_4\equiv(\mathrm{d}m/M_\odot)/(\mathrm{d}r/1000\,\mathrm{km})|_{s=4}$, 
being the
normalized mass-derivative at this location. The two parameters $\mu_4$ and
$M_4\mu_4$ can be directly linked to the mass-infall rate, $\dot M$, of the
collapsing star and the electron-type neutrino luminosity of the accreting
proto-NS, $L_{\nu_e}\propto M_\mathrm{ns}\dot M$, which play a crucial role
in the ``critical luminosity'' concept for the theoretical description of
neutrino-driven explosions as runaway phenomenon of the stalled accretion
shock. All models were evolved employing the approach of Ugliano~et~al.\
for simulating neutrino-driven explosions in spherical symmetry. The neutrino
emission of the accretion layer is approximated by a gray transport
solver, while the uncertain neutrino emission of the 1.1\,$M_\odot$ proto-NS
core is parametrized by an analytic model. The free parameters connected to
the core-boundary prescription are calibrated to reproduce the observables
of Supernova~1987A for five different progenitor models.
\end{abstract}


\section{Introduction}

Presupernova stars in the mass range above $\sim$9\,$M_\odot$ exhibit
large variations of their structure with respect to, e.g., their Fe-core
and O-core masses, their binding energies, and their density or entropy
profiles above the Fe-core  \citep{Woosley2002}. These properties
vary non-monotonically with the zero-age main-sequence (ZAMS) mass
and can differ considerably even between progenitors with only a small
difference of their ZAMS masses \citep{Sukhbold2014}.

Correspondingly, \cite{Ugliano2012} found that the properties
of neutrino-driven supernovae (SNe) like explosion energy, nickel mass,
and remnant mass change non-monotonically with the ZAMS mass.
In particular, for the
investigated grid of 101 solar-metallicity progenitors binned
in 0.2\,$M_\sun$ steps \citep{Woosley2002},
they found islands of non-exploding, black hole (BH) forming cases
down to 15\,$M_\odot$, alternating with mass-intervals of exploding
progenitors. In a few cases individual neighboring progenitors
showed opposite behavior.

\cite{Ugliano2012} used a simple, parametric model for
the contracting proto-neutron star (PNS) as a neutrino source to
trigger neutrino-driven explosions in spherically symmetric (1D)
hydrodynamic simulations, but
their basic findings were confirmed by other groups working with
semi-analytic descriptions in 1D \citep{Pejcha2015} and
approximate neutrino transport in two- and three-dimensional (2D, 3D)
hydrodynamic models \citep{Nakamura2015,Horiuchi2014}.
While \cite{OConnor2011} suggested that BH formation
requires a compactness (normalized enclosed mass-radius ratio) 
of $\xi_{2.5}>0.45$
with
\begin{equation}
\xi_{M}\equiv\frac{M/M_\odot}{R(M)/1000\,\mathrm{km}}\,,
\label{eq:compactness}
\end{equation}
\cite{Ugliano2012} obtained only explosions for $\xi_{2.5}<0.15$,
explosions or BH formation for 0.15\,$\leq\xi_{2.5}\leq$0.35,
and only BH formation for $\xi_{2.5}>0.35$, which implies a larger
fraction of BH formation cases for solar-metallicity stars.
\cite{Horiuchi2014} pointed out that a critical compactness
of $\xi_{2.5}\gtrsim 0.2$ for failed explosions is compatible with a
lack of red supergiant Type-IIP SN progenitors above $\sim$16\,$M_\odot$
\citep{Smartt2009} and with a significant excess
of the star-formation rate compared to the observed SN rate \citep{Horiuchi2011}.
\cite{Pejcha2015} showed that their parameterization ``case (a)'',
which yields
results similar to those of \cite{Ugliano2012}, is close to being optimally
compatible with a combination of several observational constraints.

How can the non-monotonicities of the explodability be understood
in terms of the pre-supernova properties and in the
context of the physics of the neutrino-driven mechanism?
Are there characteristic parameters of the pre-supernova
star that decide better about success or failure of the
explosion than a single value of the compactness or other,
similarly useful parameters like the iron-core mass or the
binding energy outside of the iron core? While all these
measures reflect trends like an enhanced tendency of BH
formation for high compactness, large iron-core mass or high
exterior binding energy, there are still many outliers that do
not obey the correlations. For example, a suitably chosen mass
$M$ of the compactness $\xi_M$ allows to correctly predict
explosions in $<$90\% of the cases \citep{Pejcha2015},
but the best choice of $M$ is merely empirical and the physical
justification of $\xi_M$ as a good diagnostics is unclear.

Here we propose a two-parameter criterion that separates successful
explosions from failures with very high reliability.
While two compactness values, e.g.\ $\xi_{1.5}$
and $\xi_{2.5}$, or the iron-core mass and the mean entropy
in some suitable mass range begin to show such a disentanglement,
we demonstrate that the normalized mass inside a dimensionless entropy per
nucleon of $s=4$,
\begin{equation}
M_4\equiv m(s=4)/M_\odot\,,
\label{eq:ms4}
\end{equation}
and the mass derivative at this location,
\begin{equation}
\mu_4\equiv\left.\frac{\mathrm{d}m/M_\odot}{\mathrm{d}r/1000\,\mathrm{km}}\right|_{s=4}\,,
\label{eq:dmdr}
\end{equation}
both determined from the pre-supernova profiles, allow to predict
the explosion behavior successfully in $\gtrsim 97$\% of all cases
and have a direct connection with the theoretical basis of the
neutrino-driven mechanism.

We briefly describe our numerical approach in Sect.~\ref{sec:numerics},
including a detailed discussion of our modeling methodology in
comparison to other approaches in the recent literature,
present our results in Sect.~\ref{sec:results}, and conclude in
Sect.~\ref{sec:conclusions}.

\section{Numerical Setup and Progenitor Models}
\label{sec:numerics}

\subsection{Modeling approach}
\label{sec:modeling}

Our basic modeling approach follows \cite{Ugliano2012} with
a number of improvements.
To trigger neutrino-driven explosions in spherically symmetric (1D)
hydrodynamic simulations, we use a schematic
model of the high-density core of the PNS as neutrino
source \citep[for details, see][]{Ugliano2012}. This analytic
description is applied to the innermost 1.1\,$M_\odot$, which are
excised from the computational domain, and it yields
time-dependent neutrino luminosities that are imposed as boundary
values at the contracting, Lagrangian inner grid boundary.
On the numerical grid, where neutrino optical depths increase from
initially $\sim$10 to finally several 1000, neutrino transport is
approximated by the gray treatment described in \cite{Scheck2006} and
\cite{Arcones2007}. This allows us to
account for the progenitor-dependent variations of the accretion
luminosity.

Our approach replaces still uncertain physics connected to the
equation of state (EoS) and neutrino opacities at high densities
by a simple, computationally efficient PNS core model. The associated
free parameters are calibrated by reproducing observational properties
of SN~1987A. We emphasize that the neutrino emission is sensitive to
the time and progenitor dependent mass accretion rate. Not only
the accretion luminosity increases for progenitors with higher mass
accretion rate of the PNS, but also the neutrino loss of the inner
core rises with the accreted mass because of compressional
work of the accretion layer on the core. Such dependences are
accounted for in our modeling of NS core and accretion.

Our numerical realization improves the treatment by
\cite{Ugliano2012} in several aspects.
We use the high-density EoS of \cite{Lattimer1991} with
a compressibility of $K = 220$\,MeV and below
$\rho=10^{11}$\,g\,cm$^{-3}$ apply an $e^\pm$, photon, and baryon
EoS \citep{Timmes2000} for nuclear statistical equilibrium
\citep[NSE;][private communication]{Kifonidis2004pc} with 16 nuclei for
$T>7\times 10^9$\,K and a 14-species alpha network (including an
additional neutron-rich tracer nucleus of iron-group material) 
at lower temperatures \citep{Muller1986}. The tracer nucleus is assumed 
to be formed in ejecta with $Y_e<0.49$ and thus tracks the ejection
of matter with neutron excess, when detailed nucleosynthesis
calculations predict little production of $^{56}$Ni
\citep{Thielemann1996}.

\begin{figure}[t]
\begin{center}
\includegraphics[width=\columnwidth]{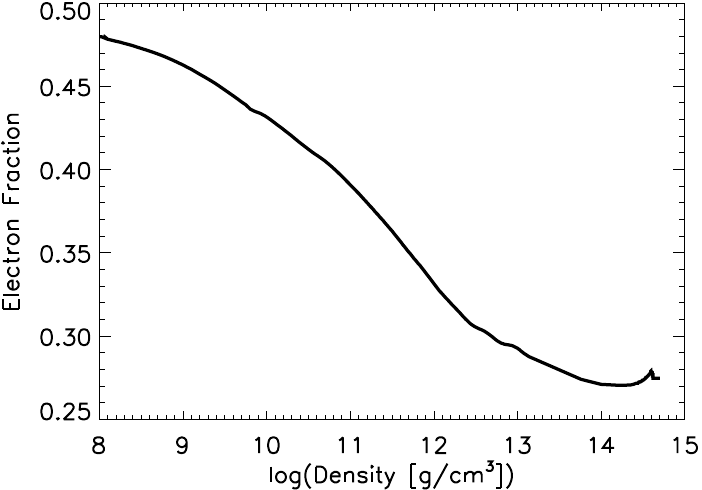}
\caption{
Electron fraction as function of density,
$Y_e(\rho)$, for modeling the deleptonization of the
collapsing stellar iron core during the infall phase until
core bounce according to the approximate treatment of
\cite{Liebendorfer2005}.
}
\label{fig:yerho}
\end{center}
\end{figure}

The network is
consistently coupled to the hydrodynamic modeling and allows
us to include the contribution from explosive nuclear burning
to the energetics of the SN explosions.
The collapse phase until core bounce is modeled with the
deleptonization scheme proposed by \cite{Liebendorfer2005},
using the $Y_e(\rho)$ trajectory of Fig.~\ref{fig:yerho} for the
evolution of the electron fraction $Y_e$ as function of density
$\rho$ \citep[][private communication]{Muller2013pc}. This
yields good overall agreement with full neutrino
transport results and allows for a very efficient computation
of large sets of post-bounce models.

\begin{deluxetable*}{cccccccccccccccc}
\tablecolumns{16}
\tabletypesize{\scriptsize}
\tablecaption{
Calibration models with explosion and remnant properties
\label{table:calibration}
}
\tablehead{
  \colhead{Calibration}                                                 &
  \colhead{\xsi\tablenotemark{a}}                                       &
  \colhead{\ensuremath{\xi_{1.75}}\tablenotemark{a}}                    &
  \colhead{\ensuremath{\xi_{2.0}}\tablenotemark{a}}                     &
  \colhead{\XSI\tablenotemark{a}}                                       &
  \colhead{$t_\text{exp}$\tablenotemark{b}}                             &
  \colhead{$E_\text{exp}$\tablenotemark{c}}                             &
  \colhead{$M_\text{ej}$\tablenotemark{d}}                              &
  \colhead{$E_\text{exp}/M_\text{ej}$}                                  &
  \colhead{$M_{{}^{56}\text{Ni}}$\tablenotemark{e} }                    &
  \colhead{$M_{\text{tracer}}$\tablenotemark{f}}                        &
  \colhead{$M_\text{ns}$\tablenotemark{g}}                              &
  \colhead{$M_\text{wind}$\tablenotemark{h}}                            &
  \colhead{$M_\text{fb}$\tablenotemark{i}}                              &
  \colhead{$t_{\nu,90}$\tablenotemark{j}}                               &
  \colhead{$E_{\nu,\text{tot}}$\tablenotemark{k}}                      \\
  \colhead{Model}                                                       &
  \colhead{}                                                            &
  \colhead{}                                                            &
  \colhead{}                                                            &
  \colhead{}                                                            &
  \colhead{[ms]}                                                        &
  \colhead{[B]}                                                         &
  \colhead{[\msun]}                                                     &
  \colhead{[B/\msun]}                                                   &
  \colhead{[\msun]}                                                     &
  \colhead{[\msun]}                                                     &
  \colhead{[\msun]}                                                     &
  \colhead{[\msun]}                                                     &
  \colhead{[\msun]}                                                     &
  \colhead{[s]}                                                         &
  \colhead{[100\,B]}
  }
\startdata
s19.8 (2002)  & 1.03   & 0.35    & 0.22   & 0.14   & 750   & 1.30  & 12.98 & 0.100      & 0.072 & 0.034 & 1.55   & 0.096  & 0.00298 & 4.27 & 3.68 \\
w15.0         & 0.34   & 0.09    & 0.03   & 0.01   & 580   & 1.41  & 13.70 & 0.103      & 0.045 & 0.046 & 1.32   & 0.088  & 0.00018 & 5.18 & 2.81 \\
w18.0         & 0.76   & 0.26    & 0.16   & 0.10   & 730   & 1.25  & 15.42 & 0.081      & 0.056 & 0.036 & 1.48   & 0.081  & 0.00310 & 4.16 & 3.32 \\
w20.0         & 0.98   & 0.35    & 0.18   & 0.06   & 620   & 1.24  & 17.81 & 0.070      & 0.063 & 0.027 & 1.56   & 0.089  & 0.00168 & 4.73 & 3.61 \\
n20.0         & 0.87   & 0.36    & 0.19   & 0.12   & 560   & 1.49  & 14.84 & 0.100      & 0.036 & 0.052 & 1.55   & 0.117  & 0.00243 & 3.97 & 3.48
\enddata
\tablenotetext{a}{Compactness evaluated for a central density of $5\times 10^{10}$\,g\,cm$^{-3}$. For w15.0 the value at core bounce is given because earlier data are not available.}
\tablenotetext{b}{Post-bounce time of onset of explosion, when shock expands beyond 500\,km (1\,B\,=\,1\,bethe\,=\,$10^{51}$\,erg).}
\tablenotetext{c}{Final explosion energy, including binding energy of preshock progenitor.}
\tablenotetext{d}{Mass ejected in the explosion.}
\tablenotetext{e}{Ejected $^{56}$Ni mass produced by explosive burning with late-time fallback taken into account.}
\tablenotetext{f}{Mass of neutron-rich tracer nucleus ejected in neutrino-driven wind material with neutron excess
(fallback is taken into account).}
\tablenotetext{g}{Final baryonic neutron-star mass including late-time fallback.}
\tablenotetext{h}{Neutrino-driven wind mass measured by mass between gain radius at $t_\mathrm{exp}$
and preliminary mass cut {\em before} fallback.}
\tablenotetext{j}{Fallback mass.}
\tablenotetext{j}{Emission time for 90\% of the radiated neutrino energy.}
\tablenotetext{k}{Total radiated neutrino energy.}
\end{deluxetable*}

\subsection{Progenitor models}
\label{sec:progenitors}

We perform collapse and explosion simulations for large progenitor
sets of different metallicities, namely: the zero-metallicity z2002
set (30 models with ZAMS masses of 11.0--40.0\,$M_\odot$),
low-metallicity ($10^{-4}$ solar) u2002 series (247 models,
11.0--75.0\,$M_\odot$), and the solar-metallicity s2002 series
(101 models, 10.8--75.0\,$M_\odot$) of \cite{Woosley2002} plus a
10.0\,$M_\odot$ progenitor \citep[][private communication]{Woosley2007pc} and a
10.2\,$M_\odot$ progenitor \citep[][private communication]{Heger2003pc};
the solar-metallicity s2014 (151 models, 15.0--30.0\,$M_\odot$)
and sh2014 series (15 models, 30.0--60.0\,$M_\odot$, no mass loss) of
\cite{Sukhbold2014}, supplemented by additional 36 models with
9.0--14.9\,$M_\odot$; the solar-metallicity s2007 series (32 models,
12.0--120.0\,$M_\odot$) of \cite{Woosley2007}; and the
n2006 series (8 models, 13.0--50.0\,$M_\odot$; \citealt{Nomoto2006}).

For the core-model parameter
calibration we choose five different progenitors, namely the (red
supergiant) model s19.8 of the s2002 series as in \cite{Ugliano2012},
and four blue supergiant pre-supernova models of SN~1987A: w15.0 (ZAMS mass of
15\,$M_\odot$; \citealt{Woosley1988}), w18.0 (18\,$M_\odot$, evolved with
rotation; \citealt{Woosley2007}), w20.0 (20\,$M_\odot$; \citealt{Woosley1997}),
and n20.0 (20\,$M_\odot$; \citealt{Shigeyama1990}). Compactness values and
explosion and remnant parameters of these models are listed in
Table~\ref{table:calibration}.

The calibration aims at producing the explosion energy and
ejected $^{56}$Ni mass of SN~1987A compatible with observations, for which
the best values are $E_\mathrm{exp}=(1.50\pm 0.12)\times 10^{51}$\,erg
\citep{Utrobin2005}, $E_\mathrm{exp}\sim 1.3\times 10^{51}$\,erg
\citep{Utrobin2011}, and
$M_\mathrm{Ni}=0.0723$--0.0772\,$M_\odot$ \citep{Utrobin2014},
but numbers reported by other authors cover a considerable range
(cf.\ \citealt{Handy2014} for a compilation). The explosion
energy that we accept for a SN1987A model in the calibration process
is guided by the ejected $^{56}$Ni mass (which fully accounts for
short-time and long-time fallback) and a ratio of $E_\mathrm{exp}$
to ejecta mass in the ballpark of estimates based on light-curve
analyses (cf.\ Table~\ref{table:calibration} for our values).

Because of the ``gentle'' acceleration of the SN shock by the
neutrino-driven mechanism \citep[also in 3D simulations, see][]{Utrobin2014},
it is difficult to produce this amount
of ejected $^{56}$Ni just by shock-induced explosive burning.
$M_\mathrm{^{56}Ni}$ in Table~\ref{table:calibration} mainly
measures this component but also contains $^{56}$Ni from proton-rich
neutrino-processed ejecta. However, also neutrino-processed ejecta
and the neutrino-driven wind with a slight neutron excess
could contribute significantly to the $^{56}$Ni production.
The electron fraction $Y_e$ of these ejecta is set by $\nu_e$
and $\bar\nu_e$ interactions and depends extremely sensitively
on the properties (luminosities and spectra) of the emitted
neutrinos, which our transport approximation
cannot reliably predict and which also depend on subtle effects
connected to multi-dimensional physics and neutrino opacities.
For these reasons we consider the $^{56}$Ni as uncertain within
the limits set by the true $^{56}$Ni yield from our network
on the low side and, in the maximal case, all tracer material added
to that. We therefore provide as possible $^{56}$Ni production
of our models the range
$M_\mathrm{^{56}Ni}\le M_\mathrm{^{56}Ni}^\mathrm{total}\le
M_\mathrm{^{56}Ni}+M_\mathrm{tracer}$.

Different from \cite{Ugliano2012} we reduce the
compression parameter of the NS core model by a linear relation
$\zeta'\propto\xi_{1.75,\mathrm{b}}$ (the value of $\xi_{1.75,\mathrm{b}}$ 
is measured at core bounce) for progenitors with $\le$13.5\,$M_\odot$,
i.e., we use the function
\begin{equation}
\zeta' = \zeta\,\left(
\frac{\xi_{1.75,\mathrm{b}}}{0.5}\right) \ \ \mathrm{for}\ \ 
M\le 13.5\,M_\odot \,,
\label{eq:scaling_zeta}
\end{equation}
with $\zeta$ being the value determined from the SN~1987A
calibration for a considered progenitor model of that supernova.
We note that the values of $\xi_{1.75,\mathrm{b}}$ are less than 0.5
for all progenitors below 13.5\,$M_\odot$ and close to 0.5 for
$M\sim 13.5\,M_\odot$, for which reason Eq.~(\ref{eq:scaling_zeta})
connects smoothly to the $\zeta$ value applied for stars above
13.5\,$M_\odot$ according to the SN~1987A calibration.

The modification of Eq.~(\ref{eq:scaling_zeta}) accounts for the reduced
burden of the small mass of the accretion layer of these stars with their
extremely low compactnesses. Such a modification allows us to reproduce
the trend to weak explosions obtained in sophisticated 2D and
3D simulations for low-mass iron-core progenitors
\citep{Janka2012a,Melson2015,Muller2015b}. 
We point out that the Crab supernova SN~1054 is considered to be
connected to the explosion of a $\sim$10\,$M_\odot$ star (e.g., 
\citealt{Nomoto1982,Smith2013}), and its explosion energy is estimated
to be up to only $\sim$10$^{50}$\,erg (e.g., \citealt{Yang2015}).
This fact lends support to the results of recent, self-consistent 1D and
multi-dimensional supernova models of $\lesssim$10\,$M_\odot$ stars
(e.g., \citealt{Kitaura2006,Fischer2010,Melson2015}), 
whose low explosion energies and low nickel production agree
with the Crab observations. 

Although as a consequence of our $\zeta$ reduction the explosion times, 
$t_\mathrm{exp}$, tend to be late for stars in the 10.5--12.5\,$M_\odot$
range (Fig.~\ref{fig:explosionprops}), this behavior also seems to be
compatible with self-consistent, multi-dimensional
simulations of stellar explosions in the 11--12\,$M_\odot$ range by
\cite{Muller2013}, \cite{Muller2015b}, and \cite{Janka2012a}, 
where these stars were
found to have a long-lasting phase of accretion and simultaneous mass
outflow after a quite inert onset of the explosion.
A more detailed discussion and justification of our modified treatment of
low-ZAMS mass cases will be provided in Sect.~\ref{sec:calibration}
and can also be found in a follow-up paper by \cite{Sukhbold2015},
where the values of all PNS core-model parameters are
tabulated for all calibrations.

It must be emphasized that in the context of the present work the
detailed treatment of the low-mass stars is not overly important.
These stars usually explode fairly easily, independent of the 
treatment of the PNS core model with the original or with
our revised calibration. Therefore these stars lie far away from
the boundary curve that separates exploding from non-exploding models
and whose determination will be our main goal in Sect.~\ref{sec:results}.
For this reason exactly the same separation line is obtained when the 
core-model parameter values from the SN~1987A calibration are applied to
all stars.

In Table~\ref{table:calibration} and the rest of our paper,
time-dependent structural parameters of the stars (like compactness
values, $\mu_4$ of Eq.~\ref{eq:dmdr}, the iron-core mass $M_\mathrm{Fe}$)
are measured when the stars possess a central density of
$5\times 10^{10}$\,g\,cm$^{-3}$, unless otherwise stated. This choice
of reference density defines a clear standard for the comparison of
stellar profiles of different progenitors \cite[cf.\ Appendix~A of][]{Buras2006}.
Different from the moment of core bounce,
which was used in other works, our reference density has the
advantage to be still close to the initial state of the pre-collapse
models provided by stellar evolution modeling and therefore to yield
values of the structural parameters that are more similar to those
of the pre-collapse progenitor data. Our calibration model w15.0,
however, must be treated as an exception. Because pre-collapse
profiles of this model are not available any more, all structural
quantities for this case are given (roughly) at core bounce.

\begin{figure*}
\begin{center}
\includegraphics[width=\textwidth]{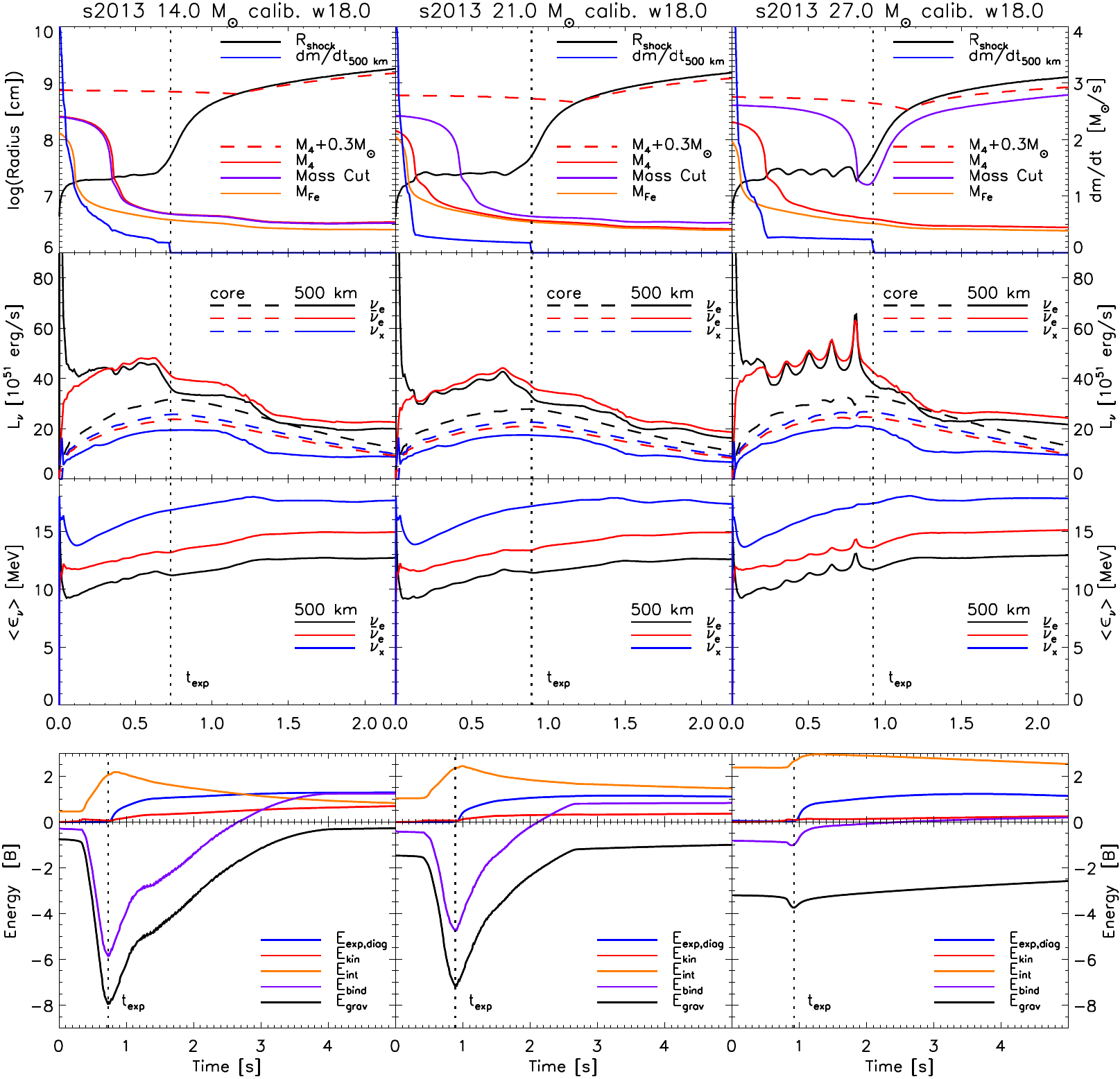}
\caption{
Models s14.0 ({\em left}), s21.0 ({\em middle}), and s27.0 ({\em right})
of the s2014 progenitor series as exemplary cases of successful explosions
with the w18.0 calibration. The {\em top panels} display as functions of
post-bounce time the radius of
the outgoing shock (black line), the mass accretion rate measured at
500\,km (blue line; scale on the right side), and the radii of iron core
(orange), $M_4=m(s=4)$ (red), $M_4+0.3$ (red dashed) and trajectory
of the final mass cut (after completion of fallback; purple).
The {\em second panels from top} show the time evolution of the luminosities
of $\nu_e$, $\bar\nu_e$ and a single species of heavy-lepton neutrinos $\nu_x$
as labelled in the plot, measured at 500\,km (solid lines) and at the inner
grid boundary (dashed lines). The {\em third panels from top} show the mean
energies of all neutrino kinds as radiated at 500\,km. The vertical dotted
lines indicate the onset time of the explosion as the moment when the
outgoing shock passes the radius of 500\,km.
The {\em bottom panels} provide the time evolution of the diagnostic energy
of the explosion (integrated energy of all postshock zones with positive
total energy; blue line). Also shown are the kinetic energy (red),
gravitational energy (black), and internal energy (orange) as integrals
over the whole, final SN ejecta between the final mass cut (after fallback)
on the one side and the stellar surface on the other. The total (binding)
energy (purple) as the sum of these energies ultimately converges to the
diagnostic energy and both of these energies asymptote to the final
explosion energy. While this convergence is essentially reached after
$\sim$4\,s in the case of s14.0, the expansion of shocked matter in the s21.0
model and thus the energy evolution is slowed down at $\sim$2.7\,s by the
high densities in the stellar core. The s27.0 model becomes gravitationally
unbound (i.e., the total binding energy becomes positive) even more slowly
because of the very massive stellar core. The convergence of total energy and
diagnostic energy takes tens of seconds in this case.
}
\label{fig:expl-examples}
\end{center}
\end{figure*}

\subsection{Methodology and theory of neutrino-driven explosions}
\label{sec:methodology}

\subsubsection{Status of ``ab initio'' supernova modeling}
\label{sec:abinitio}

``Ab initio'', fully self-consistent simulations of stellar core
collapse with state-of-the-art treatment of microphysics and
neutrino transport do not lead to explosions in spherical symmetry
except for stars with O-Ne-Mg and Fe-cores near the low-mass
end of SN progenitors 
\citep{Kitaura2006,Janka2008,Janka2012a,Fischer2010,Melson2015}.
2D simulations
in the recent past have produced successful explosions and
underline the fundamental importance of multidimensional effects, but
the true meaning of these results with respect to
the neutrino-driven mechanism is not finally clear and the
current situation is diffuse and contradictive.

On the one hand, some
of the 2D explosions set in relatively late and might remain on the
weak side \citep[e.g.,][]{Marek2009,Suwa2012,Suwa2014,Muller2012a,Muller2012b,
Muller2013,Muller2014,Muller2015b,Hanke2014},
although such apprehension is
speculative because not all simulations could be continued until
the explosion energy had saturated \citep{Muller2015b}. 
On the other hand, the Oak Ridge
group obtained explosions much earlier after bounce with shock
evolutions being astonishingly similar for 12, 15, 20, and 25\,$M_\odot$
stars and explosion energies fairly compatible with observations
\citep{Bruenn2013,Bruenn2014}. In contrast, \cite{Dolence2015}
did not find any successes in 2D simulations of the
same progenitors but they used a different treatment of gravity,
hydrodynamics,
equation of state, neutrino transport, and neutrino opacities.
The exact reasons for the different findings will have to be
clarified by detailed tests and comparisons.
The situation is even more diffuse because current 3D simulations
agree in showing slower explosions compared to 2D calculations or
even no explosions \citep[e.g.,][]{Hanke2012,Hanke2013,Tamborra2014,Mezzacappa2015,Couch2013a,Couch2014,Takiwaki2014}
although some studies have proclaimed the
opposite behavior \citep{Nordhaus2010,Burrows2013,Dolence2013}.
So far only a few recent 3D calculations with highly refined neutrino
treatment have obtained successful shock revival by the neutrino-driven
mechanism \citep{Melson2015,Melson2015b,Lentz2015}. Interestingly, 
the 3D simulation of a low-mass (9.6\,$M_\odot$) progenitor with
detailed neutrino physics, whose explosion energy approached its
saturation level, was found to explode more energetically
in 3D than in 2D \citep{Melson2015}. This result is in line with
a 2D-3D comparison in the 11--12\,$M_\odot$ range conducted by
\cite{Muller2015b}. In both studies accretion downflows and the 
re-ejection of neutrino-heated matter were observed to be different
in 2D and 3D because of geometry dependent differences of the 
Kelvin-Helmholtz instability and flow fragmentation. The 3D models
therefore suggest that explosions in 2D are massively affected
by the assumption of rotational symmetry around the polar grid
axis and by an inverse turbulent energy cascade, which tends to 
amplify energy on the largest possible scales \citep{Hanke2012} and
also produces numerical artifacts in the post-explosion 
accretion phase of the neutron star \citep{Muller2015b}. It must
therefore be suspected that the early onset of explosions and
the extremely unipolar or bipolar
deformations along the symmetry axis obtained in many 2D models
could be artifacts of the imposed symmetry constraints.

\subsubsection{Modeling recipes in recent literature}

Before 3D modeling will have become a routine task and
results will have converged, neutrino-driven explosions
of large sets of progenitor stars can be explored for their
observational implications only by referring to simplified
modeling approaches. Several different recipes
have been introduced for this recently. \cite{Ugliano2012}
used an analytic PNS core-cooling model in connection with
a neutrino transport approximation in 1D hydrodynamic explosion
simulations (as briefly summarized in Sect.~\ref{sec:numerics}), thus
improving the simpler, time-dependent boundary neutrino luminosity
prescribed by previous users of the simulation code
\citep{Kifonidis2006,Scheck2004,Scheck2006,Scheck2008,Arcones2007,Arcones2011}
and the even simpler neutrino light-bulb treatment (without any
transport approximation) applied by \cite{Janka1996} and
\cite{Kifonidis2003}. \cite{OConnor2011} resorted
to a scaling parameter $f_\mathrm{heat}$ to artificially enhance
the neutrino heating by charged-current processes behind the stalled
shock in 1D hydrodynamic models with approximate neutrino
treatment. \cite{Nakamura2015} performed an extensive set
of 2D simulations with simplified neutrino transport despite the
grains of salt mentioned in Sect.~\ref{sec:abinitio}
\citep[see also][for cautioning against the 2D results]{Horiuchi2014}.
\cite{Pejcha2015} applied a semi-analytic model to
determine the onset times of the explosions, using neutrino
luminosities from 1D calculations of accreting PNSs, and estimated
explosion properties by analytic arguments. \cite{Suwa2014}
also performed 2D simulations and suggested analytic approximations
for describing diffusion and accretion components of the neutrino
luminosities from PNSs and a free-fall treatment for the collapse
of the overlying stellar layers.
\cite{Perego2015} invented a method
they named ``PUSH'', which they applied to trigger explosions
artificially in their general relativistic, 1D hydrodynamic
core-collapse and PNS formation modeling with sophisticated neutrino
transport. PUSH gradually switches on and off
additional neutrino heating of chosen strength during a chosen
period of time. This procedure is assumed to mimic the effects of
multi-D hydrodynamics in the postshock region.
The extra heating is coupled to the {\em heavy-lepton
neutrino} emission from the PNS.

All of these recipes contain larger sets of parameters and degrees
of freedom, which are either varied in exploring different cases
\citep[e.g.,][]{Pejcha2015} or are adjusted by comparison
to more complete models \cite[e.g.,][]{Suwa2014} or by
reproducing observational benchmarks like those set by
SN~1987A \citep{Ugliano2012,Perego2015}.
The models of \cite{Ugliano2012} and those in the present paper 
assume that the explosion trigger is tightly coupled to the
physics that reflects the main differences between different progenitor
stars, namely to the post-bounce accretion history of the collapsing
stellar core and the corresponding accretion luminosities of $\nu_e$ 
and $\bar\nu_e$. It will have to be seen whether this important aspect
of the models remains being supported by future developments towards a 
more complete understanding of the physics of the central engine that
powers the explosion in the context of the neutrino-driven mechanism.

\begin{figure*}
\begin{center}
\includegraphics[width=0.975\textwidth]{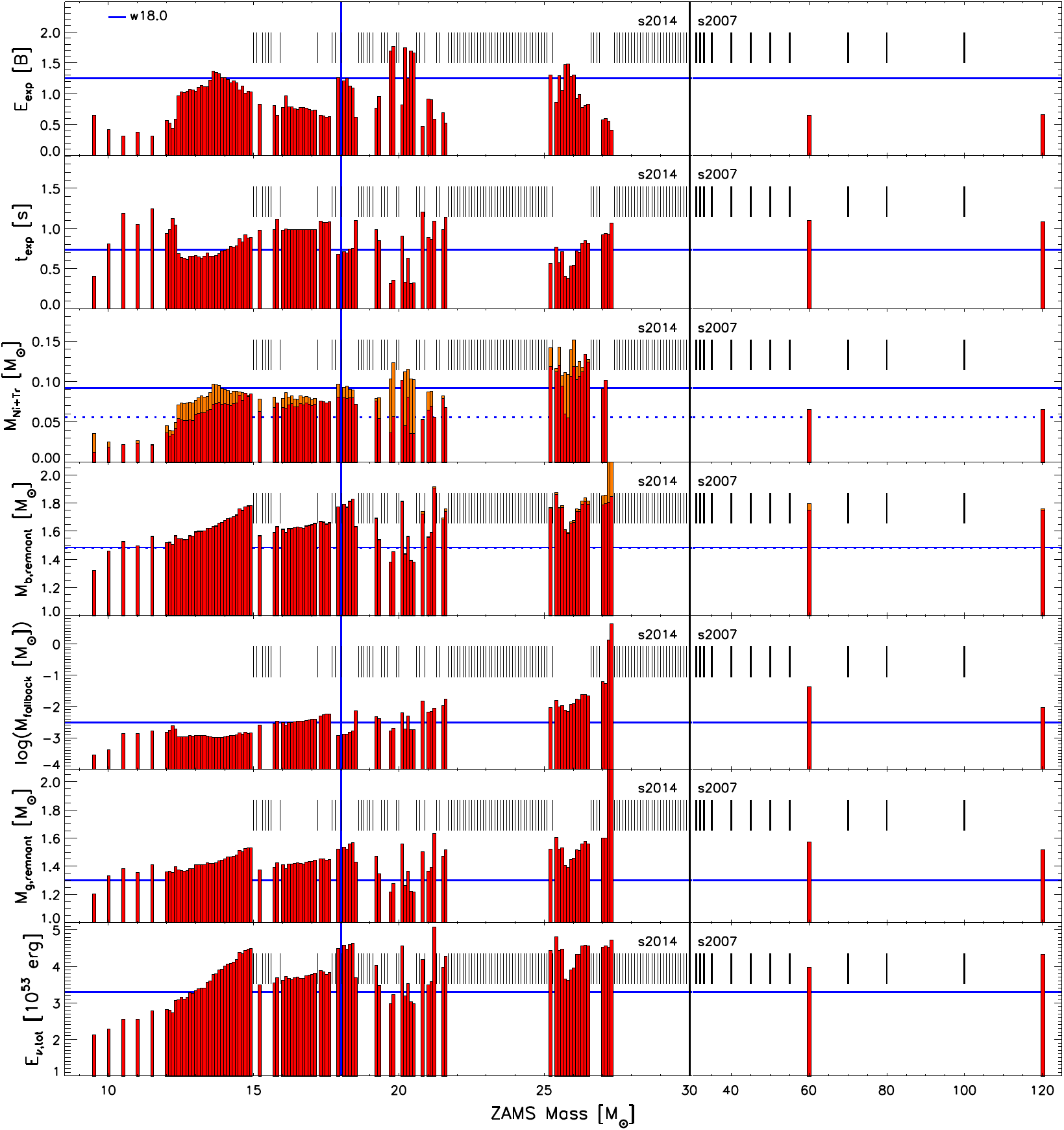}
\caption{Explosion properties for models of the s2014 series,
the supplementary low-mass progenitors with
$M_\mathrm{ZAMS}<15\,M_\odot$, and models with
$M_\mathrm{ZAMS}>30\,M_\odot$ from the s2007 series,
exploded with the w18.0 calibration. A black vertical line marks the boundary
between the two progenitor sets.
The panels show the final explosion energies, $E_\mathrm{exp}$ ({\em top};
1\,B\,=\,1\,bethe\,=\,$10^{51}$\,erg),
times of the onset of the explosion, $t_\mathrm{exp}$ (defined as the
moment when the shock expands beyond 500\,km; {\em second from top}),
masses of ejected explosively produced $^{56}$Ni (red bars) and tracer
element (orange bars; {\em third from top}), baryonic remnant masses with
fallback masses indicated by orange sections of the bars ({\em fourth from top}),
fallback masses (plotted logarithmically; {\em fifth from top}),
gravitational remnant masses (Eq.~(\ref{eq:gravmass}); {\em sixth from top}),
and total energies radiated in neutrinos, $E_{\nu,\mathrm{tot}}$
({\em bottom}). Mass and parameter values of the calibration models
are indicated by vertical and horizontal blue lines, respectively,
with the dashed horizontal blue line in the middle panel giving the
$^{56}$Ni mass and the solid horizontal blue line the sum of $^{56}$Ni
and tracer. Non-exploding cases are marked by short vertical black
bars in the upper half of each panel.
}
\label{fig:explosionprops}
\end{center}
\end{figure*}
\begin{figure*}
\begin{center}
\includegraphics[width=0.495\textwidth]{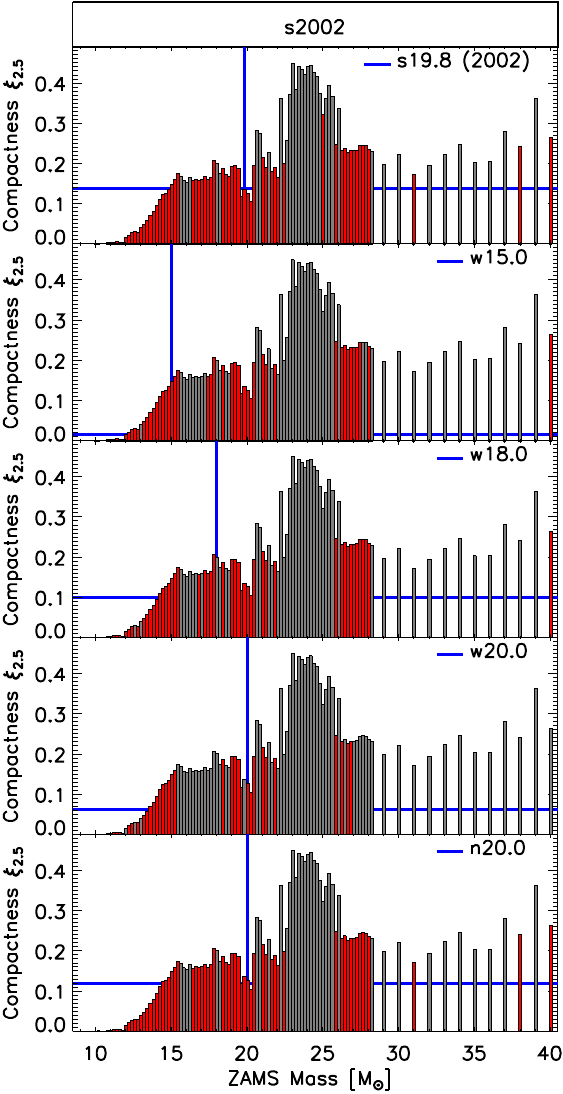}\hskip1.0pt
\includegraphics[width=0.495\textwidth]{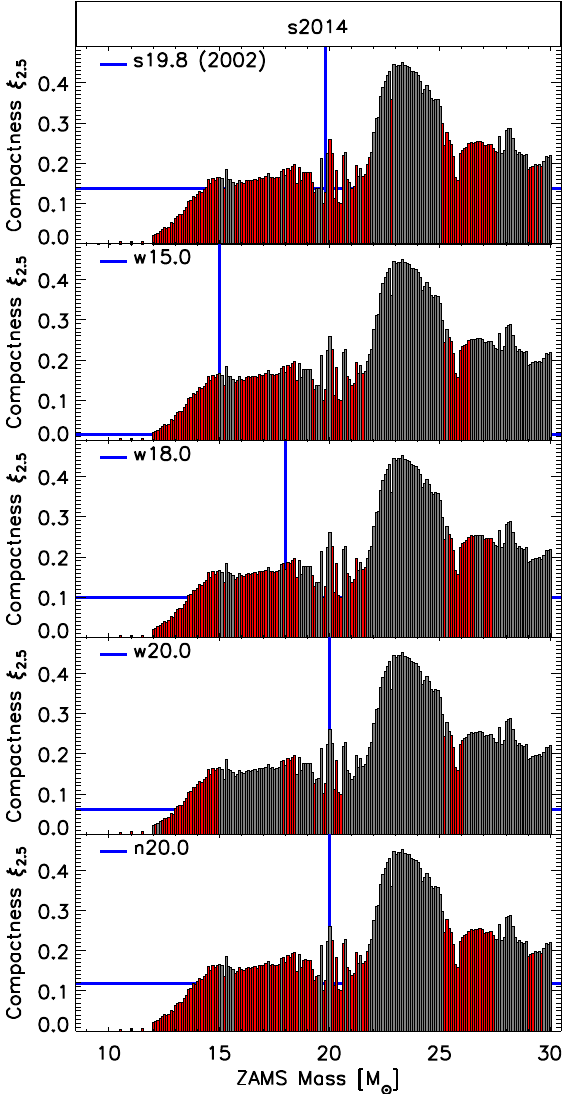}
\caption{
Compactness $\xi_{2.5}$ versus ZAMS mass for the s2002 ({\em left}) 
and s2014
progenitor series (plus $M_\mathrm{ZAMS}<15\,M_\odot$ models)
with exploding (red bars) and non-exploding (gray bars) cases
for all calibrations. The s2002 results for the s19.8 calibration agree well
with \cite{Ugliano2012}. Blue vertical and horizontal lines indicate
the values of the progenitors used for the calibrations.
}
\label{fig:explosions}
\end{center}
\end{figure*}

\subsubsection{Motivation of modeling assumptions of this work}

The analytic NS-core model introduced by \cite{Ugliano2012}
in combination with their approximate transport solver for
treating the accretion component of the neutrino luminosity as
well as neutrino cooling and heating between PNS and shock,
is an attempt to realize the tight coupling of accretion
behavior and explodability in close similarity to what is found in
current 2D simulations \citep[e.g. those of][]{Marek2009,Muller2012a,Muller2012b,
Muller2013,Muller2014}.
Since 1D models with elaborate neutrino physics and a fully
self-consistent calculation of PNS cooling miss the critical
condition for explosions by far, it is {\em not the goal} of
\cite{Ugliano2012} to closely reproduce the neutrino
emission properties of such more sophisticated calculations.
Rather than that it is the goal
to approximate the {\em combined effects of
neutrino heating and multi-dimensional postshock hydrodynamics}
by a simple and computationally efficient neutrino source model,
which allows for the fast processing of large progenitor sets including
the long-time evolution of the SN explosion to determine also the
shock breakout and fallback evolution.

Free parameters in the NS core
model and the prescribed contraction behavior of the inner grid boundary
are calibrated by matching basic observational features (explosion
energy, $^{56}$Ni yield, total release of neutrino energy) of SN~1987A.
This is intended to ensure that the overall properties of the
neutrino-source model are anchored on empirical ground.
Of course, the setting of the parameter values cannot be unambiguous
when only a few elements of a single observed SN are used for
deriving constraints. However, the approximate nature of the
neutrino source treatment as a whole does not require the perfectly
accurate description of each individual model component in order
to still contain the essence of the physics of the system
like important feedback effects between accretion and outflows
and neutrinos, which govern the
progenitor-dependent variations of explodability and SN
properties. A reasonable interplay of the different components is
more relevant than a most sophisticated representation of any
single aspect of the neutrino source model.

In detail, the basic features of our 1D realization of the
neutrino-driven mechanism along the
lines of \cite{Ugliano2012} are the following:
\begin{itemize}
\item
The possibility of an explosion is coupled closely to the
progenitor-dependent strength and evolution of the post-bounce
accretion. This is achieved not only by taking into account
the accretion luminosity through the approximate neutrino
transport scheme but also through the response of the PNS-core
to the presence of a hot accretion mantle. The evolution of the
latter is explicitly followed in our hydrodynamic simulations,
which track
the accumulation of the collapsing stellar matter around the
inner PNS core. The existence of the mantle layer enters the
analytic core model in terms of the parameter $m_\mathrm{acc}$
for the mass of this layer and the corresponding accretion
rate $\dot m_\mathrm{acc}$.
\item
The inner 1.1\,$M_\odot$ core of the PNS is cut out and replaced
by a contracting inner grid boundary and a corresponding boundary
condition in our model. This inner core is considered to be
the supranuclear high-density region of the nascent NS,
whose detailed physics is still subject to considerable
uncertainties. This region is replaced by an analytic description,
whose parameters $\Gamma$, $R_\mathrm{c}(t)$, and $n$
\citep[see][]{Ugliano2012} are set to the same values for all
stars. This makes sense because the supranuclear phase is highly
incompressible, for which reason it can be expected that the
volume of the core is not largely different during the explosion
phase for different PNS masses. Moreover, the
neutrino diffusion time scale out of this core
is seconds, which implies that its neutrino emission is of
secondary importance during the shorter post-bounce
phase when the explosion develops.
Despite its simplicity, our core treatment still
includes progenitor and accretion dependent variations
through the mass $m_\mathrm{acc}$ of the hot accretion mantle
of the PNS and the mass accretion rate $\dot m_\mathrm{acc}$,
whose influence on the inner core is accounted for in Eqs.~(1)--(4)
of \cite{Ugliano2012} for describing the energy evolution of the
core model.
\item
The onset of the explosion is considerably delayed (typically
between several 100\,ms and about a second) with a slow (instead
of abrupt) rise of the explosion energy during the subsequent
shock acceleration phase, when an intense neutrino-driven wind
ejects matter and delivers power to the explosion. Neutrino
heating cannot deposit the explosion energy impulsively, because
the ejected matter needs to absorb enough energy from neutrinos
to be accelerated outwards. The rate of energy input to the
explosion is therefore limited by the rate at which matter can
be channeled trough the heating region. A long-lasting
period (hundreds of milliseconds to more than a second)
of increasing energy is characteristic of neutrino-driven
explosions (Fig.~\ref{fig:expl-examples}) and is observed as
gradual growth of the explosion energy also in 2D explosion
models, e.g., by \cite{Scheck2006} and \cite{Bruenn2014}.
To achieve this behavior in our 1D models the
core-neutrino source needs to
keep up high neutrino luminosities for a more extended period
of time than found in fully self-consistent SN simulations, where
the rapid decline of the mass-accretion rate at the surface of
the iron core and at the interface of silicon and silicon-enriched
oxygen layers leads to a strong decrease of the accretion
luminosity. The longer period of high neutrino emission is
compensated by a somewhat underestimated early post-bounce
neutrino luminosity (Fig.~\ref{fig:expl-examples})
in order to satisfy the energy constraints
set by the total gravitational binding energy of the forming NS.
\item
The time scale and duration of the growth of the explosion energy
in multi-D models of neutrino-driven SNe are connected to an extended
period of continued accretion and simultaneous shock expansion
that follows {\em after} the revival of the stalled shock
\citep[see][]{Marek2009}. Persistent accretion thereby
ensures the maintenance of a significant accretion luminosity, while
partial re-ejection of accreted and neutrino-heated matter boosts
the explosion energy.
In our 1D simulations the physics of such a two-component flow
cannot be accurately accounted for. In order to approximate the
consequences of this truly multi-dimensional phase, our 1D models
are constructed with important two properties: On the one hand
they refer to a high level of the PNS-core luminosity for about
one second. On the other hand they
are set up to possess a more extended accretion phase that
precedes the onset of the delayed explosion before the intense
neutrino-driven wind pumps energy into the explosion. The power
and mass loss in this wind are overestimated compared to
sophisticated neutrino-cooling simulations of PNSs. However,
this overestimation of the wind strength has its justification:
The early wind is supposed to mimic the mass ejection that is
fed in the multi-dimensional case
by the inflow and partial re-ejection of matter falling
towards the gain radius during the episode of simultaneous
accretion and shock expansion. The enhanced wind mass
counterbalances the extra mass accretion by the PNS during the
long phase before the shock acceleration is launched, and this
enhanced wind mass is of crucial importance to carry the energy
of the neutrino-powered blast.
\end{itemize}

Figure~\ref{fig:expl-examples} shows the post-bounce evolution of
the stalled SN shock, the onset of the explosion, energy evolution,
and the time evolution of the neutrino emission properties for
three representative progenitors, namely s14, s21, and s27 of the
s2014 series, which explode successfully with the w18.0 calibration.
The shock stagnation at a radius of approximately 200\,km lasts between
$\sim$700\,ms and 900\,ms and can exhibit the well-known oscillatory
expansion and contraction phases, which signal proximity to the
explosion \citep[see e.g.,][]{Buras2006,Murphy2008,Fernandez2012}.
The explosion sets in shortly after $M_4$
(Eq.~\ref{eq:ms4}) has fallen through the shock and well before
the mass shell corresponding to $M_4+0.3\,$ has collapsed.
High neutrino luminosities are maintained by high mass accretion
rates and, after the onset of the explosion, by
a contribution from the core emission (dashed lines in
the luminosity panel of Fig.~\ref{fig:expl-examples})
that grows until roughly one second. The
current models underestimate the surface luminosity of heavy-lepton
neutrinos compared to more sophisticated simulations because of
the chosen modest contraction of the inner boundary of the
computational grid (which leads to underestimated temperatures
in the accretion layer of the PNS) and
because neutrino-pair production by nucleon-nucleon
bremsstrahlung is not taken into account. We did not upgrade
our treatment in this respect because $\nu_\mu$ and $\nu_\tau$
are not of immediate relevance for our study since the explosion
hinges exclusively on the heating by $\nu_e$ and $\bar\nu_e$.

Replacing the inner core of the PNS by a contracting inner
boundary of the computational mesh introduces a number of
free parameters, whose settings allow one to achieve the
desired accretion and neutrino-emission behavior as detailed
above. On the one hand, our model contains parameters
for the prescription of the contraction of the grid boundary,
on the other hand there are parameters for the
simple high-density core model (cf.\  \citealt{Ugliano2012}
 and references therein).
While the core-model parameters
($\Gamma$, $\zeta$, $R_\mathrm{c}(t)$, $n$)
regulate the neutrino-emission evolution of the excised,
high-density core of the PNS, the prescribed grid-boundary
radius, $R_\mathrm{ib}(t)$, governs the settling
of the hot accretion mantle of the PNS. Because
of partially compensating influences and dependences, not all
of these parameters have a sensitive impact on the outcome
of our study. Again, more relevant than a highly accurate
description of individual components of the modeling is a
reasonable reproduction of the overall properties of the
accretion and neutrino emission history of the stalled SN
shock and mass accumulating PNS. For example, the moderate
increase of the mean neutrino energies with time and their
regular hierarchy ($\langle\epsilon_{\nu_e}\rangle <
\langle\epsilon_{\bar\nu_e}\rangle <
\langle\epsilon_{\nu_x}\rangle$; Fig.~\ref{fig:expl-examples})
are not compatible with the most sophisticated
current models \citep[see, e.g.,]{Marek2009,Muller2014}.
They reflect our choice of a less
extreme contraction of the 1.1\,$M_\odot$ shell than found
in simulations with soft nuclear equations of state for
the core matter, where the PNS contracts more strongly and
its accretion mantle heats up to higher temperatures at
later post-bounce times (compare \citealt{Scheck2006} and see
the discussion by \citealt{Pejcha2015}). Our choice is
motivated solely by numerical reasons (because of less stringent
time-step constraints), but it has no immediate drawbacks for
our systematic exploration of explosion conditions in large
progenitor sets. Since neutrino-energy deposition
depends on $L_\nu\langle\epsilon_{\nu}^2\rangle$, the
underestimated mean neutrino energies at late times can be
compensated by higher neutrino luminosities $L_\nu$ of the
PNS core.

The neutrino emission from the PNS-core region is
parametrized in accordance with basic physics constraints.
This means that the total loss of electron-lepton number is
compatible with the typical neutronization of the inner
1.1\,$M_\odot$
core, whose release of gravitational binding energy satisfies
energy conservation and virial theorem \citep[see][]{Ugliano2012}.
Correspondingly chosen boundary luminosities
therefore ensure a basically realistic deleptonization and
cooling evolution of the PNS as a whole and of the accretion
mantle in particular,
where much of the inner-boundary fluxes are absorbed and
reprocessed. Again, a proper representation of
progenitor-dependent variations requires a reasonable description
of the overall system behavior but does not need a very high
sophistication of all individual components of the system.

\subsubsection{Calibration for low-ZAMS mass range}
\label{sec:calibration}

Agreement with the constraints from SN~1987A employed in our work
(i.e., the observed explosion energy, $^{56}$Ni mass, total neutrino
energy loss, and the duration of the neutrino signal) can be
achieved with different sets of values of the PNS core-model 
parameters. Using only one observed SN case the parameter set is
underconstrained and the choice of suitable values is ambiguous.
It is therefore not guaranteed that the calibration works
equally well in the whole mass range of investigated stellar models.

In particular stars in the low ZAMS-mass
regime ($M_\mathrm{ZAMS}\lesssim 12$--13\,$M_\odot$) possess
properties that are distinctly different from those of the
adopted SN~1987A progenitors and in the mass
neighborhood of these progenitors. Stars with 
$M_\mathrm{ZAMS}\lesssim 12$--13\,$M_\odot$ are characterized
by very small values of compactness (Eq.~\ref{eq:compactness}),
binding energy outside of the iron core and outside of $M_4$
(Eq.~\ref{eq:ms4}), and mass derivative $\mu_4$ (Eq.~\ref{eq:dmdr}).
The progenitor of SN~1054 giving birth to the Crab remnant
is considered to belong to this mass range, more specifically to
have been a star with mass around 10\,$M_\odot$
\citep{Nomoto1982,Smith2013,Tominaga2013}. Because of their
structural similarities and distinctive differences compared
to more massive stars, \cite{Sukhbold2015} call progenitors
below roughly 12--13\,$M_\odot$ ``Crab-like'' in contrast to
stars above this mass limit, which they term ``SN~1987-like''.

Stars below $\sim$10\,$M_\odot$ were found to explode easily in 
self-consistent, sophisticated 1D, 2D and 3D simulations
\citep{Kitaura2006,Janka2008,Fischer2010,Wanajo2011,Janka2012a,Melson2015}
with low energies (less than or around $10^{50}\,\mathrm{erg} = 0.1$\,B)
and little nickel production ($<\,$0.01\,$M_\odot$), in agreement with
observational properties concluded from detailed analyses of the 
Crab remnant (e.g., \citealt{Yang2015}). We therefore consider
the results of these state-of-the-art SN models together with the
empirical constraints for the Crab supernova as important 
benchmark that should be reproduced by our approximate 1D modeling of 
neutrino-powered explosions.

The results of \cite{Ugliano2012} revealed a problem
in this respect, because they showed far more energetic
explosions of stars in the low-mass domain than expected
on grounds of the sophisticated simulations and from 
observations of Crab. Obviously, the
neutrino-source calibration used by \cite{Ugliano2012} is
not appropriate to reproduce ``realistic'' explosion
conditions in stars with very dilute shells
around the iron core. Instead, it leads to an overestimated power
of the neutrino-driven wind and therefore overestimated explosion
energies. In particular, the strong and energetic wind is in conflict 
with the short period of simultaneous postshock accretion and mass
ejection after the onset of the explosion in $\lesssim$10\,$M_\odot$
stars. Since the mass accretion rate is low and the duration of the
accretion phase is limited by the fast shock expansion, the energetic
importance of this phase is diminished by the small mass that is 
channelled through the neutrino-heating layer in convective flows
\citep{Kitaura2006,Janka2008,Wanajo2011,Janka2012a,Melson2015}. 
In order to account for these features found in the most refined
simulations of low-mass stellar explosions, the neutrino-driven wind
power of our parametric models has to be reduced.

We realize such a reduction of the wind power by decreasing
the parameter
$\zeta$, which scales the compression work exerted on the inner
(excised) core of the PNS by the overlying accretion mantle
\citep[cf.\ Eqs.~1--4 in][]{Ugliano2012},
in proportionality to the compactness parameter $\xi_{1.75,\mathrm{b}}$,
which drops strongly for low-mass progenitors 
(see Eq.~\ref{eq:scaling_zeta}). This procedure
can be justified by the much lighter accretion layers of such
stars, which implies less compression of the PNS core by
the outer weight. Such a modification reduces the neutrino
emission of the high-density core and therefore the mass outflow
in the early neutrino-driven wind. As a consequence, the explosion
energy falls off towards the low-mass end of the investigated
progenitor sets. This can be seen in the upper panel of
Fig.~\ref{fig:explosionprops}, which should be compared to
the upper left panel of Fig.~5 in \cite{Ugliano2012}.

The $\zeta$ scaling of Eq.~(\ref{eq:scaling_zeta})
is introduced as a quick fix in the course of this work and
is a fairly ad hoc measure to cure the problem of overestimated 
explosion energies for low-mass SN progenitors. 
In \cite{Sukhbold2015} a different approach is taken, in which the
final value of the core-radius parameter of the one-zone model 
describing the supernuclear PNS interior as
neutrino source (see \citealt{Ugliano2012})
is modified. This procedure can directly be
motivated by the contraction of the PNS found in 
self-consistent cooling simulations with microphysical high-density
equations of state and detailed neutrino transport. Mathematically,
the modification of the core radius has a similar effect on the
core-neutrino emission as the $\zeta$ scaling employed here.
While we refer the reader to \cite{Sukhbold2015} for details,
we emphasize that the consequences for the overall explosion
behavior of the low-mass progenitors is very similar for both
the $\zeta$ reduction and the core-radius adjustment applied
by \cite{Sukhbold2015}. They lead to considerably lower 
explosion energies for $M_\mathrm{ZAMS}\lesssim 12$\,$M_\odot$
stars and a further drop of the explosion energies below 
$\sim$9.5\,$M_\odot$.

As a drawback of this modification, the explosions of
some of the low-mass progenitors between $\sim$10.5\,$M_\odot$
and $\sim$12.5\,$M_\odot$ set in rather late
($>$1\,s\,p.b., cf.\ Fig.~\ref{fig:explosionprops})\footnote{In
extremely rare cases one may even obtain failed explosions.}. 
This, however,
is basically compatible with the tendency of relatively slow
shock expansion and late explosions
that are also found in sophisticated multi-dimensional simulations
of such stars, which, in addition, reveal long-lasting phases 
of simultaneous accretion and mass ejection after the onset of the
explosion (\citealt{Muller2013,Muller2015b}).
This extended accretion phase has only moderate consequences
for the estimated remnant masses, because the mass accretion
rate of these progenitors reaches a low level of
$\lesssim$0.05--0.1\,$M_\odot$\,s$^{-1}$ after a few 100\,ms post
bounce, and some or even most of the accreted mass is re-ejected
in the neutrino-driven wind.

Besides providing information on the explosion energies and
the onset times of the explosion (defined by the time the
outgoing shock reaches 500\,km), Fig.~\ref{fig:explosionprops}
also displays for exploding models the ejected masses of $^{56}$Ni 
and iron-group tracer element, the baryonic and gravitational
remnant masses, the fallback masses, and the total energies
radiated by neutrinos. Overall, these results exhibit features
very similar to those discussed in detail by \cite{Ugliano2012}
for a different progenitor series and a different
calibration model. We point out that the fallback masses in
the low-mass range of progenitors were overestimated by
\cite{Ugliano2012} due to an error in the analysis (more
discussion will follow in Sect.~\ref{sec:systematics};
an erratum on this aspect is in preparation.)

\begin{figure}
\begin{center}
\includegraphics[width=\columnwidth]{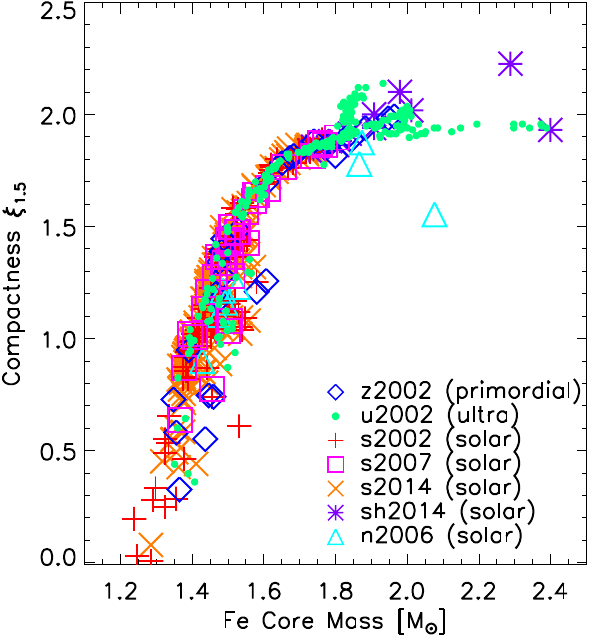}
\caption{
Correlation of iron-core mass, $M_\mathrm{Fe}$,
and compactness $\xi_{1.5}$ for
the investigated models of all progenitor series. Note that
$M_\mathrm{Fe}$ is taken from the pre-collapse model while
$\xi_{1.5}$ is evaluated for a central density
of $5\times 10^{10}$\,g\,cm$^{-3}$.
}
\label{fig:mfe-xicorr}
\end{center}
\end{figure}

\begin{figure*}
\begin{center}
\includegraphics[width=0.98\textwidth]{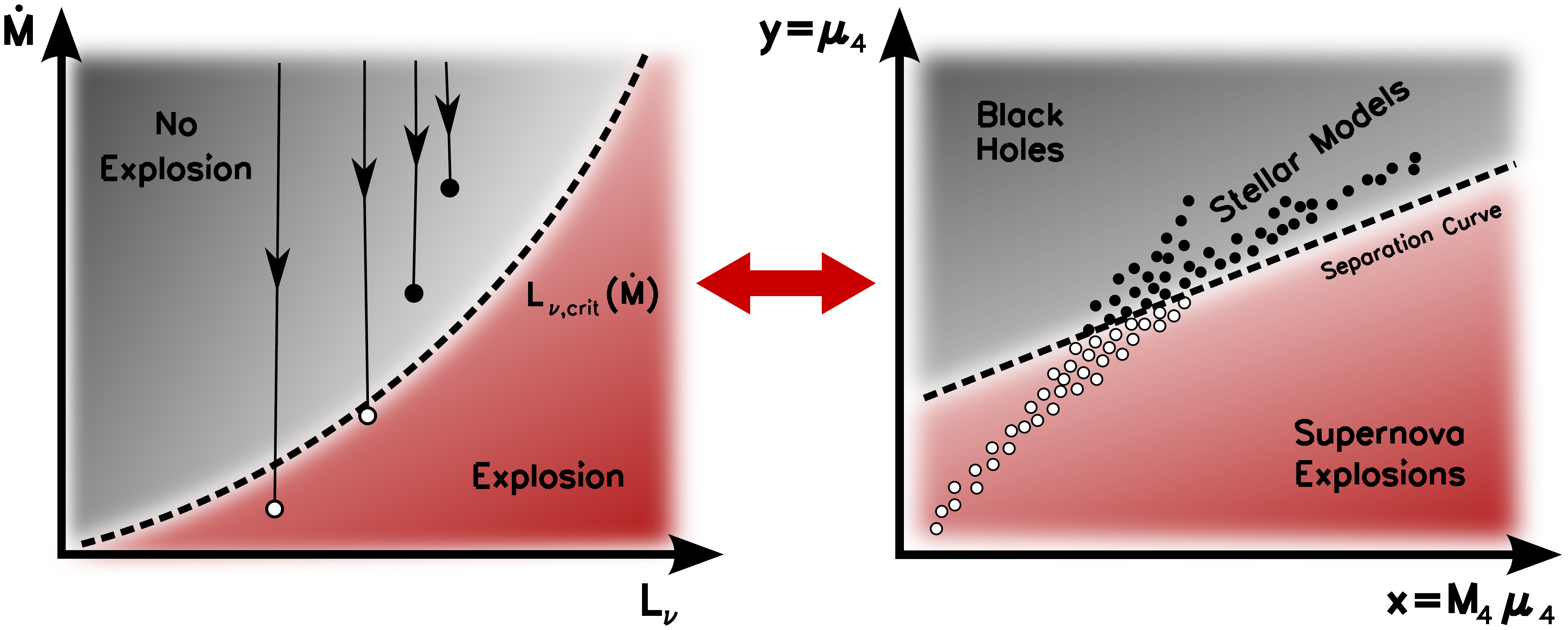}
\caption{
Correspondence of $L_\nu$-$\dot M$ plane with critical neutrino luminosity
$L_{\nu,\mathrm{crit}}(\dot M)$ ({\em left}) and $x$-$y$ plane with separation
curve $y_\mathrm{sep}(x)$ ({\em right}). In the left plot post-bounce evolution
paths of successfully exploding models (white circles) and non-exploding models
(black circles) are schematically indicated, corresponding to white and black
circles for pre-collapse models in the right plot.
Evolution paths of successful models cross the critical line at some point
and the accretion ends after the explosion has taken off. In contrast,
the tracks of failing cases never reach the critical conditions for
launching the runaway expansion of the shock. The symbols in the left plot
mark the ``optimal point'' relative to the critical curve that can be
reached, corresponding to the stellar conditions described by the
parameters $(M_4\mu_4,\mu_4)$ at the $s=4$ location, which seems
decisive for the success or failure of the explosion of a progenitor,
because the accretion rate drops strongly outside.
}
\label{fig:cartoon}
\end{center}
\end{figure*}

\section{Results}
\label{sec:results}

\subsection{One and two-parameter classifications}
\label{sec:twopars}

Figure~\ref{fig:explosions} shows $\xi_{2.5}$ versus ZAMS mass with
BH formation cases indicated by gray and explosions by red bars
for the s2002 and s2014 series and all calibrations.
The irregular pattern found by
\cite{Ugliano2012} for the s2002 progenitors is reproduced
and appears similarly in the s2014 set. High compactness $\xi_{2.5}$
exhibits a tendency to correlate with BHs. But also
other parameters reflect this trend, for example
$\xi_{1.5}$, the iron-core mass $M_\mathrm{Fe}$
(defined as the core where $\sum_{\mathrm{\{}i|A_i>46\mathrm{\}}}X_i>0.5$
for nuclei with mass numbers $A_i$ and mass fractions $X_i$),
and the enclosed mass at the bottom
of the O-burning shell. All three of them are tightly correlated,
see Fig.~\ref{fig:mfe-xicorr} as well as Fig.~4
of \cite{Ugliano2012}. Also high values of the
binding energy $E_\mathrm{b}(m>M_\mathrm{Fe})$ outside of
$M_\mathrm{Fe}$ signals a tendency for BH formation, because
this energy correlates with $\xi_{2.5}$ \citep[cf.\ Fig.~4 in][]{Ugliano2012}.
However, for none of these single
parameters a sharp boundary value exists that discriminates
between explosions and non-explosions. For all such choices of a
parameter, the BH formation limit tends to vary (non-monotonically)
with $M_\mathrm{ZAMS}$ and in a broad interval of values either
explosion or BH formation can happen. \cite{Pejcha2015}
tried to optimize the choice of $M$ for $\xi_M$, but even their
best case achieved only 88\% of correct predictions. Since in the
cases of $\xi_{2.5}$ and
$E_\mathrm{b}(m>M_\mathrm{Fe})$, for example, the threshold value
for BH formation tends to grow with higher ZAMS mass, one may
hypothesize that a second parameter could improve the predictions.

Placing the progenitors in a two-parameter space spanned by
$\xi_{1.5}$ and $\xi_{2.5}$ or, equally good, $M_\mathrm{Fe}$
and $\xi_{2.5}$, begins to show a cleaner separation of successful
and failed explosions: SNe are obtained for small values of
$\xi_{2.5}$, whereas BHs are formed for high values of $\xi_{2.5}$,
but the value of this threshold increases with $\xi_{1.5}$ and
$M_\mathrm{Fe}$. For given $\xi_{1.5}$ (or $M_\mathrm{Fe}$) there
is a value of $\xi_{2.5}$ above which only BHs are formed.
However, there is still a broad overlap region of mixed cases.

This beginning separation can be understood in view of the
theoretical background of the neutrino-driven mechanism, where the
expansion of the SN shock is obstructed by the ram pressure of infalling
stellar-core matter and shock expansion is pushed by neutrino-energy
deposition behind the shock. For neutrino luminosities above a
critical threshold $L_{\nu,\mathrm{crit}}(\dot M)$, which depends on
the mass-accretion rate $\dot M$ of the shock, shock runaway and
explosion are triggered by neutrino heating
\cite[see Fig.~\ref{fig:cartoon}, left panel, and, e.g.,][]{Burrows1993,
Janka2001,Murphy2008,Nordhaus2010,Hanke2012,Fernandez2012,Pejcha2012,
Janka2012b,Muller2015}.
$M_\mathrm{Fe}$ (or $\xi_{1.5}$)
can be considered as a measure of the mass $M_\mathrm{ns}$
of the PNS as accretor,
which determines the strength of the gravitational potential
and the size of the neutrino luminosities. Such a dependence can
be concluded from the proportionality $L_\nu\propto R_\nu^2T_\nu^4$,
where $R_\nu$ is the largely
progenitor-independent neutrinosphere radius and the
neutrinospheric temperature $T_\nu$ increases roughly linearly with
$M_\mathrm{ns}$ \citep[see][]{Muller2014}. On the other
hand, the long-time mass-accretion rate of the PNS grows
with $\xi_{2.5}$, which is higher for denser stellar cores.
For each PNS mass explosions become impossible above a certain
value of $\dot M$ or $\xi_{2.5}$.

\subsection{Two-parameter classification based on the theoretical
concept of the neutrino-driven mechanism}

If the initial mass cut at the onset of the explosion
develops at an enclosed mass $M=m(r)$ of the progenitor, we
can choose $M=m(r)$ as a suitable proxy of the initial PNS mass,
$M_\mathrm{ns}$. A rough measure of the
mass-accretion rate $\dot M$ by the stalled shock around the onset
of the explosion is then given by the mass-gradient
$m'(r)\equiv\mathrm{d}m(r)/\mathrm{d}r=4\pi r^2\rho(r)$ at the
corresponding radius $r$. This is the case
because $m'(r)$ can be directly linked
to the free-fall accretion rate of matter collapsing into the shock
from initial radius $r$ according to
\begin{eqnarray}
\dot M=\frac{\mathrm{d}m}{\mathrm{d}t_\mathrm{ff}}&=&
\frac{\mathrm{d}m(r)}{\mathrm{d}r}
\left(\frac{\mathrm{d}t_\mathrm{ff}(r)}{\mathrm{d}r}\right)^{-1}\nonumber\\
&=&\frac{2m'(r)}{t_\mathrm{ff}[(3/r)-m'(r)/m(r)]}
\approx\frac{2}{3}\,\frac{r}{t_\mathrm{ff}}\,m'(r)\,,
\label{eq:macc}
\end{eqnarray}
where $t_\mathrm{ff}=\sqrt{r^3/[Gm(r)]}$ is the free-fall timescale
\citep{Suwa2014} and the last, approximate equality is justified
by the fact that $(m'/m)^{-1}=\mathrm{d}r/\mathrm{d}\ln(m)\gg r$
outside of the dense stellar core.

Following the critical-luminosity concept now points the way to
further improvements towards a classification scheme of explosion
conditions: The $L_\nu$-$\dot M$ dependence of the neutrino-driven
mechanism suggests that the explodability of the progenitors may be
classified by the parameters $M=m(r)$ and $\dot M\propto m'(r)$,
because the accretion luminosity $L_\nu^\mathrm{acc}\propto
GM_\mathrm{ns}\dot M/R_\mathrm{ns}\propto Mm'(r)$ accounts for
a major fraction of the neutrino luminosity of the PNS at the
time of shock revival \citep{Muller2014,Muller2015}, and, in
particular, it is the part of the neutrino emission that reflects
the main progenitor dependence. It is important to note that the
time-evolving NS and neutrinospheric radii, $R_\mathrm{ns}\sim R_\nu$,
are nearly the same for different progenitors and only weakly 
time dependent when the explosions take place rather late after bounce.
This is true for our simulations
(where $t_\mathrm{exp}\gtrsim 0.5$\,s with few exceptions; 
Fig.~\ref{fig:explosionprops}) as well as for self-consistent,
sophisticated models (see, e.g., Fig.~3 in \citealt{Muller2014}).
In both cases the spread of the NS radii and their evolution between 
$\sim$0.4\,s and 1\,s after bounce accounts for less than 25\%
variation around an average value of all investigated models.
In Sect.~\ref{sec:nsradius}
we will come back to this argument and give reasons 
why the NS radius has little influence on the results discussed
in this work. Moreover, the neutrino loss from the 
low-entropy, degenerate PNS core, whose properties are determined
by the incompressibility of supranuclear matter, should exhibit a
progenitor dependence mostly through the different weight of the
surrounding accretion mantle, whose growth depends on $\dot M$. Such
a connection is expressed by the terms depending on $m_\mathrm{acc}$
and $\dot m_\mathrm{acc}$ in the
neutrino luminosity of the high-density PNS core in Eq.~(4)
of \cite{Ugliano2012}. We therefore hypothesize, and demonstrate
below, a correspondence of the $L_\nu$-$\dot M$ space and the
$Mm'$-$m'$ parameter plane and expect that the
critical luminosity curve $L_{\nu,\mathrm{crit}}(\dot M)$
maps to a curve separating BH formation and successful
explosions in the $Mm'$-$m'$ plane\footnote{Since the shock revival
is determined by neutrino heating, which depends on
$L_\nu\langle\epsilon_\nu^2\rangle$, and since the average squared
neutrino energy $\langle\epsilon_\nu^2\rangle\propto T_\nu^2\propto
M_\mathrm{ns}^2$ increases with time and PNS mass
(\citealt{Muller2014}),
\cite{Muller2015} discuss the critical condition for shock revival
in terms of $L_\nu\langle\epsilon_\nu^2\rangle$ as a function of
$M_\mathrm{ns}\dot M$. This suggests that an alternative choice of
parameters could be $M^3m'$ and $Mm'$ instead of $Mm'$ and $m'$,
respectively. Our results demonstrate that the basic physics is
already captured by the $Mm'$-$m'$ dependence.}.

In our simulations neutrino-driven explosions set in
around the time or shortly after the moment
when infalling matter arriving at the shock possesses an
entropy $s\sim 4$. We therefore choose $M_4$
(Eq.~\ref{eq:ms4}) as our proxy of the PNS mass, 
$M_4\propto M_\mathrm{ns}$, and $\mu_4\equiv
m'(r)[M_\odot/(1000\,\mathrm{km})]^{-1}|_{s=4}$ (Eq.~\ref{eq:dmdr}) 
as corresponding measure of
the mass-accretion rate at this time, $\mu_4 \propto\dot M$.
The product $Mm'$ is therefore repesented by $M_4\mu_4$.
Tests showed that replacing
$M_4$ by the iron-core mass, $M_\mathrm{Fe}$, is similarly good
and yields results of nearly the same quality in the
analysis following below (which points to an underlying
correlation between $M_4$ and $M_\mathrm{Fe}$). In practice,
we evaluate Eq.~(\ref{eq:dmdr}) for $\mu_4$ by the average
mass-gradient of the progenitor just outside of $s=4$ according to
\begin{align}
\mu_4
&\equiv\left.\frac{\Delta m/\msun}{\Delta r/1000\,\mathrm{km}}
\right|_{s=4} \nonumber \\
&=\frac{(\msf+\Delta m/\msun)-\msf}{\left[r(\msf+
\Delta m/\msun)-r(s=4)\right]/1000\,\mathrm{km}}\,,
\label{eq:dm_dr2}
\end{align}
with $\Delta m=0.3\,M_\odot$ yielding optimal results according to
tests with varied mass intervals $\Delta m$.
With the parameters $M_4$ and $\mu_4$ picked, our imagined
mapping between critical conditions in the $L_\nu$-$\dot M$ 
and $Mm'$-$m'$ spaces transforms into such a mapping relation 
between the $L_\nu$-$\dot M$ and $M_4\mu_4$-$\mu_4$ planes
as illustrated by Fig.~\ref{fig:cartoon}.

\begin{figure*}
\begin{center}
\includegraphics[width=\textwidth]{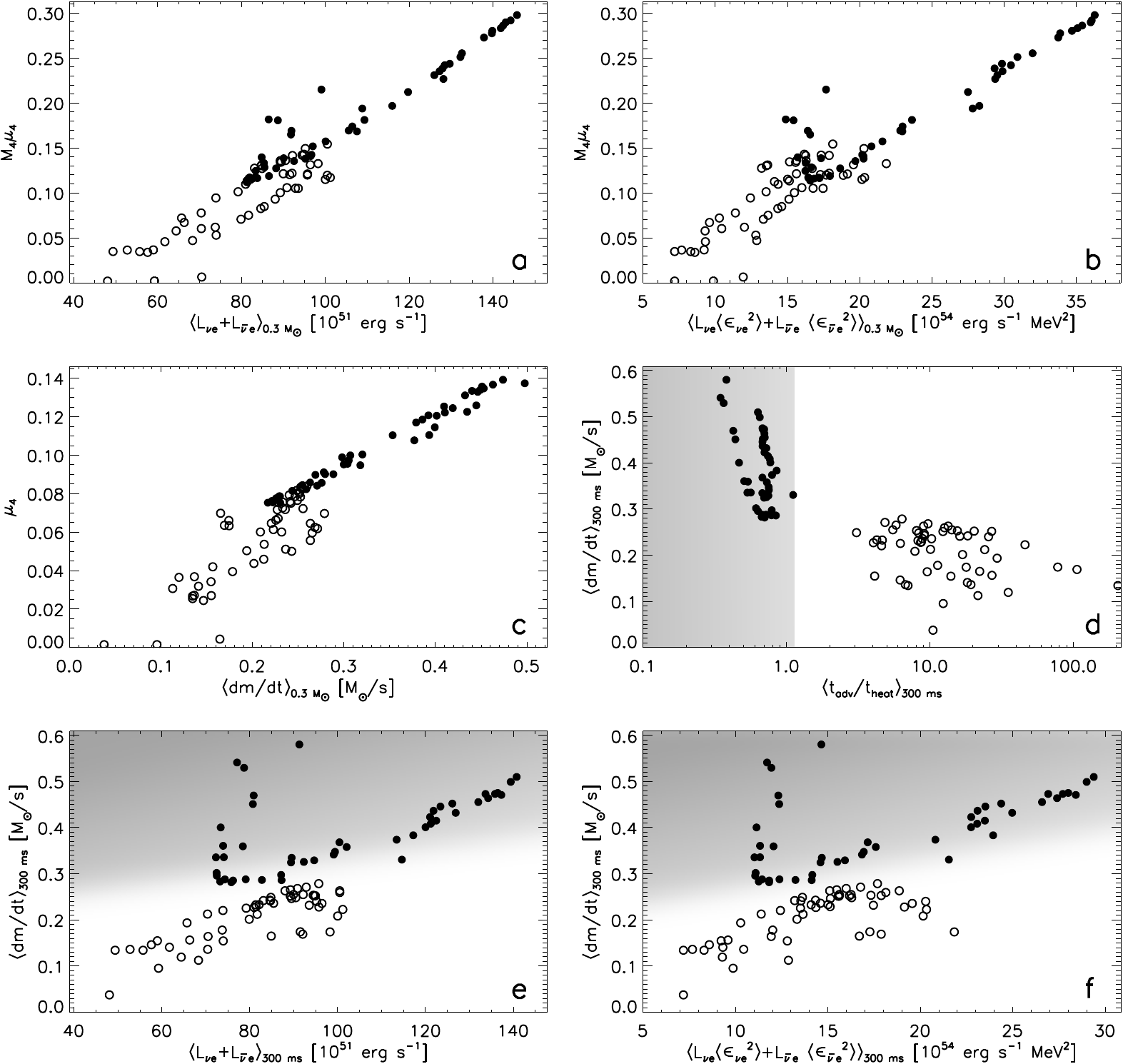}
\caption{End points of the postbounce evolution of exploding (open circles)
and non-exploding models (filled circles) in planes spanned by various
pairs of parameters. The data correspond to results of the s2002 model
series with w18.0 calibration. All symbols represent time-averaged
locations because of strong temporal variations of the postshock 
accretion layer.
{\em Panels a--c} display correlations of our dimensionless progenitor
parameters $\mu_4$ and $M_4\mu_4$ with time-averaged values of the preshock
mass accretion rate, $\dot M = \mathrm{d}m/\mathrm{d}t$,
and the characteristic neutrino-emission properties, respectively,
as obtained in our simulations and measured at 500\,km. The abscissas
of {\em panels a and b} show the summed luminosities of $\nu_e$ and
$\bar\nu_e$, $L_{\nu_e}+L_{\bar\nu_e}$,
and the summed products of luminosities and
mean squared energies,
$L_{\nu_e}\langle\epsilon_{\nu_e}^2\rangle +
L_{\bar\nu_e}\langle\epsilon_{\bar\nu_e}^2\rangle$,
of both neutrino species, respectively. For the
exploding models the time averaging is performed from
the arrival of the $s=4$ interface at the shock until the explosion sets
in (defined by the shock radius reaching 500\,km), whereas the averages
for non-exploding models cover the time from the
$s=4$ interface passing the shock until 0.3\,$M_\odot$ of overlying
material have been accreted by the shock.
{\em Panel d} displays the separation of exploding and
non-exploding models in the plane spanned by the mass-accretion
rate and the ratio of advection to heating time scale.
{\em Panels e and f} demonstrate this separation in the planes spanned by
$\dot M$ and $L_{\nu_e}+L_{\bar\nu_e}$ or
$L_{\nu_e}\langle\epsilon_{\nu_e}^2\rangle +
L_{\bar\nu_e}\langle\epsilon_{\bar\nu_e}^2\rangle$, respectively.
The time averages of the quantities in {\em panels d--f} are computed
from the passage of the $s=4$ interface through the shock until 300\,ms
later for non-exploding models or until the onset of the explosion
otherwise. Gray shading in {\em panels d--f} indicates the regions where
explosions fail. (No exact boundary curves are determined for the 
cases of {\em panels e and f}.)
}
\label{fig:correlations}
\end{center}
\end{figure*}

Figure~\ref{fig:correlations} demonstrates the strong correlation
of the mass accretion rate $\dot M=\mathrm{d}m/\mathrm{d}t$
with the parameter $\mu_4$ as given by Eq.~(\ref{eq:dm_dr2}) 
(panel~c) as well as the tight correlations between $M_4\mu_4$
and the sum of $\nu_e$ and $\bar\nu_e$ luminosities (panel~a)
and the summed product of the luminosities and mean squared energies
of $\nu_e$ and $\bar\nu_e$ (panel~b). It is important to note that
the non-stationarity of the conditions requires us to average
the quantities plotted on the abscissas over time from the moment 
when the $s=4$ interface passes through the shock until the models
either explode (i.e., the shock radius expands beyond 500\,km; open
circles) or until the mass shell $(M_4+0.3)\,M_\odot$ has fallen
through the shock, which sets an endpoint to the time interval 
within which explosions are obtained (non-exploding cases marked
by filled circles). The time averaging is needed not only because
of evolutionary changes of the preshock mass-accretion rate 
(as determined by the progenitor structure) and corresponding 
evolutionary trends of
the emitted neutrino luminosities and mean energies. The averaging
is necessary, in particular, because the majority of our models
develops large-amplitude shock oscillations after the accretion 
of the $s=4$ interface, which leads to quasi-periodic variations
of the neutrino emission properties with more or less pronounced,
growing amplitudes (see the examples in Fig.~\ref{fig:expl-examples}).
Panel~d demonstrates that exploding models (open circles) exceed 
a value of unity for the ratio of advection time 
scale, $t_\mathrm{adv}$, to heating time scale, $t_\mathrm{heat}$,
in the gain layer, which was considered as a useful critical
threshold for diagnosing explosions in many previous works   
(e.g., \citealt{Janka1998,Thompson2000,Janka2001b,Thompson2005,Buras2006,Marek2009,Muller2012b,Muller2015,Fernandez2012}).
The exploding models also populate the region towards low mass-accretion
rates (as visible in panels a--c, too), which confirms our observation 
reported in Sect.~\ref{sec:twopars}. 
In contrast, non-exploding models cluster,
clearly separated, in the left, upper area of panel~d, where 
$t_\mathrm{adv}/t_\mathrm{heat}\lesssim 1$ and the mass-accretion
rate tends to be higher. For the calculation
of the time scales we follow the definitions previously used by, 
e.g., \cite{Buras2006}, \cite{Marek2009}, \cite{Muller2012b},
\cite{Muller2015}:
\begin{eqnarray}
t_\mathrm{heat}&=&\left(\int_{R_\mathrm{g}}^{R_\mathrm{s}} (e+\Phi)
\rho \, \mathrm{d}V \right)\left(\int_{R_\mathrm{g}}^{R_\mathrm{s}}  
\dot q_\nu \rho \, \mathrm{d}V\right)^{-1} \,,\label{eq:t_heat} \\
t_\mathrm{adv} &=& \int_{R_\mathrm{g}}^{R_\mathrm{s}}
\frac{1}{\left|v_r\right|} \, \mathrm{d}r  \,.
\label{eq:t_adv}
\end{eqnarray}
Here, the volume and radius integrals are performed over the 
gain layer between gain radius $R_\mathrm{g}$ and shock radius
$R_\mathrm{s}$. $e$ is the sum of the specific kinetic
and internal energies, $\Phi$ the (Newtonian) gravitational potential,
$\rho$ the density, $\dot q_\nu$ the net heating rate per unit of mass,
and $v_r$ the velocity of the flow. 
Again, because of the variations of the diagnostic quantities
associated with the time evolution of the collapsing star and the 
oscillations of the gain layer,
the mass-accretion rate and time-scale ratio are time-averaged
between the moment when the $s=4$ interface passes the shock until
either 300\,ms later or until the model explodes (shock radius 
exceeding 500\,km)\footnote{We tested intervals ranging from 100\,ms
to 600\,ms and observed the same trends for all choices.}.
Panels~e (and f) lend support to the concept of a critical threshold
luminosity in the $L_\nu$-$\dot M$ (and 
$L_\nu\langle\epsilon_\nu^2\rangle$-$\dot M$) space mentioned above.
The two plots show a separation of exploding (open circles) and 
non-exploding (filled circles) models in a plane spanned by the
time-averaged values of the preshock mass accretion rate on the
one hand and
the sum of $\nu_e$ and $\bar\nu_e$ luminosities (panel~e) or
the summed product of $\nu_e$ and $\bar\nu_e$ luminosities times 
their mean squared energies on the other hand (panel~f). 
Again the same time-averaging
procedure as for panel~d is applied. After the $s=4$ progenitor
shell passes the shock, the time-averaged conditions of the 
exploding models reach the lower halfs of these panels, whereas the
time-averaged properties of the non-exploding models define the
positions of these unsuccessful explosions in the upper halfs. A 
separation appears that can be imagined to resemble the critical
luminosity curve $L_{\nu,\mathrm{crit}}(\dot M)$ sketched in the
left panel of Fig.~\ref{fig:cartoon}.

Because of the strong time dependence of the postshock conditions
and of the neutrino emission during the phase of dynamical shock
expansion and contraction after the accretion of the $s=4$ interface,
it is very difficult to exactly determine the critical luminosity
curve that captures the physics of our exploding 1D simulations. 
Although we do not consider such an effort as hopeless if the 
governing parameters are carefully taken into account (cf., e.g.,
the discussions in \citealt{Pejcha2015,Muller2015}) and their time 
variations are suitably averaged, we think that the clear separation 
of successful and failed explosions visible in panels~e and f of
Fig.~\ref{fig:correlations} provides proper and sufficient support 
for the notion of such a critical curve (or, more general:
condition) in the $L_{\nu,\mathrm{crit}}$-$\dot M$ 
space. Figure~\ref{fig:cartoon} illustrates our imagined relation
between the evolution tracks of collapsing
stellar cores that explode or fail to explode in this
$L_\nu$-$\dot M$ space on the left side and the locations of
SN-producing and BH-forming progenitors in the $M_4\mu_4$-$\mu_4$
plane on the right side. The sketched evolution paths are guided by
our results in panels~e and f of Fig.~\ref{fig:correlations}.
In the following section we will demonstrate that exploding and
non-exploding simulations indeed separate in the $M_4\mu_4$-$\mu_4$
parameter space.

\subsection{Separation line of exploding and non-exploding progenitors}
\label{sec:sepline}

\begin{figure*}
\begin{center}
\includegraphics[width=.495\textwidth]{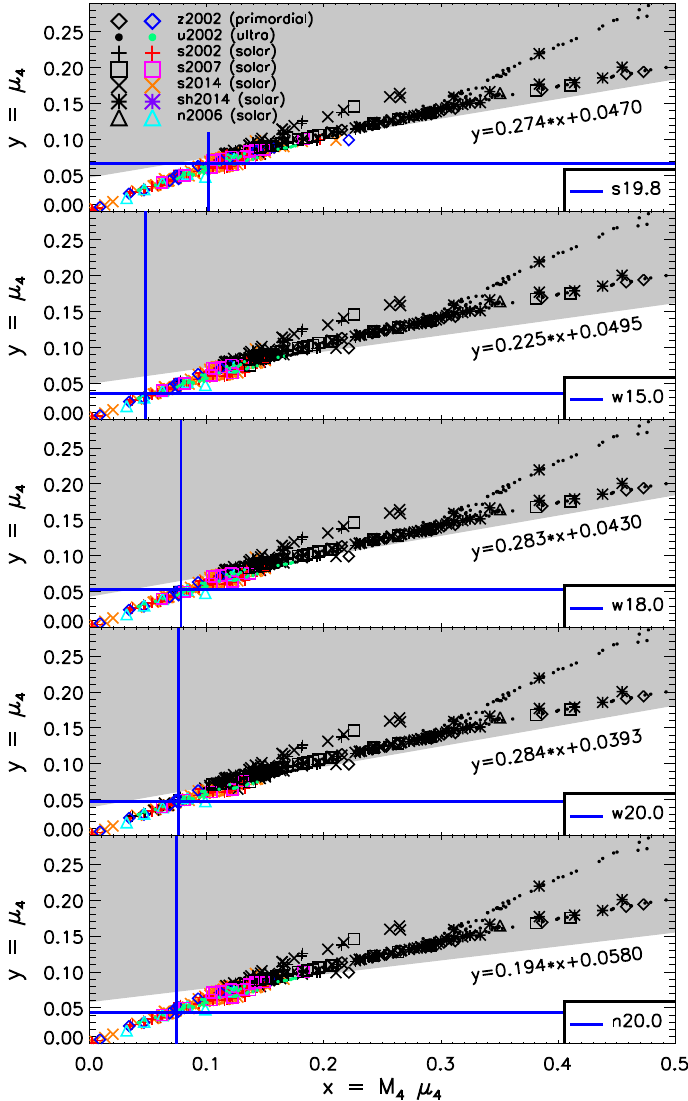}\hskip4.0pt
\includegraphics[width=.495\textwidth]{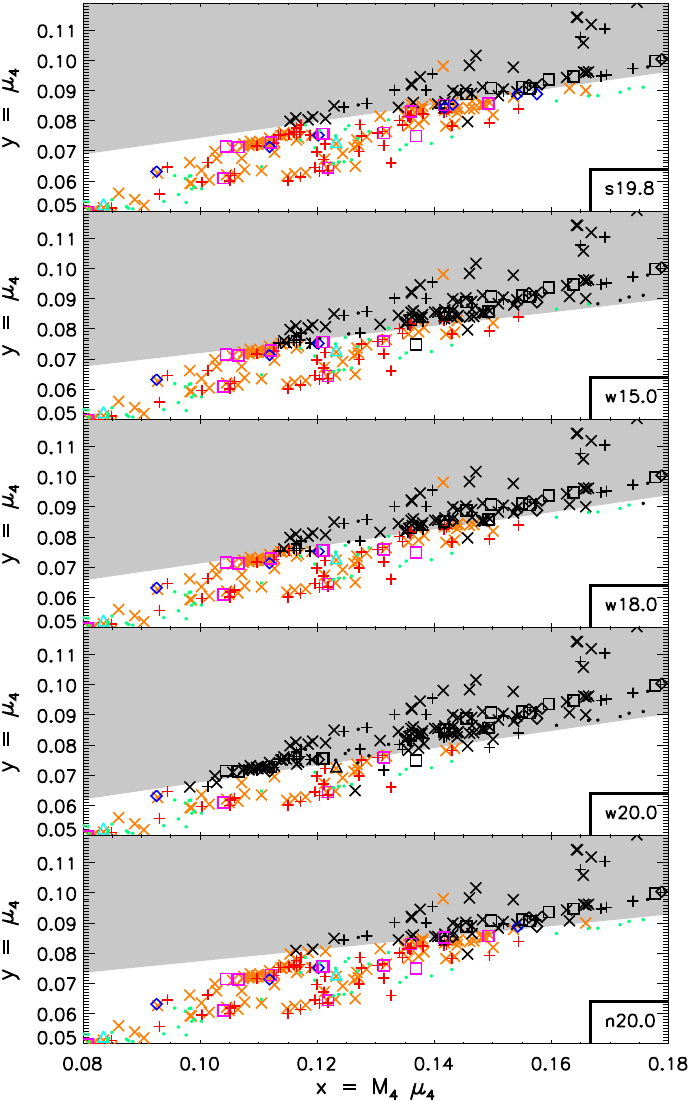}
\caption{
Separation curves between BH formation (gray region, black symbols) and
SN explosions (white region, colored symbols) for all calibrations in the
plane of parameters $x=M_4\mu_4$ and $y=\mu_4$ (zooms in {\em right} panels).
Note that the left panels do not show roughly two dozen BH-forming models of
the u2002 series, which populate the $x$-range between 0.5 and 0.62 and are
off the displayed scale.
Different symbols and colors correspond to the different
progenitor sets. The locations of the calibration models are also indicated
in the left panels by crossing blue lines.
}
\label{fig:criterion}
\end{center}
\end{figure*}

The existence of a separation line between BH forming and
SN producing progenitors in the $M_4\mu_4$-$\mu_4$ plane is 
demonstrated by Fig.~\ref{fig:criterion}, which shows the 
positions of the progenitors for all investigated model series in 
this two-dimensional parameter space.
For all five calibrations successful
explosions are marked by colored symbols, whereas BH formation is
indicated by black symbols. The regions of failed explosions are
underlaid by gray. They are bounded by straight lines with fit
functions as indicated in the panels of Fig.~\ref{fig:criterion},
\begin{equation}
y_\mathrm{sep}(x)=k_1\cdot x+k_2\,,
\label{eq:sepcurve}
\end{equation}
where $x\equiv M_4\mu_4$ and $y\equiv\mu_4$ are dimensionless
variables with $M_4$ in solar masses and $\mu_4$ computed by
Eq.~(\ref{eq:dm_dr2}). The values of
the dimensionless coefficients $k_1$ and $k_2$
as listed in Table~\ref{table:separation}
are determined by minimizing the numbers of outliers.

\begin{deluxetable}{cccccc}
\tablecolumns{3}
\tabletypesize{\scriptsize}
\tablecaption{
Bh-sn separation curves for all calibration models
\label{table:separation}
}
\tablehead{
  \colhead{Calibration Model}                                           &
  \colhead{$k_1$\tablenotemark{a}}                                      &
  \colhead{$k_2$\tablenotemark{a}}                                      &
  \colhead{\msf\tablenotemark{b}}                                       &
  \colhead{\scc\tablenotemark{b}}                                       &
  \colhead{\msf\scc\tablenotemark{b}}
  }
\startdata
s19.8 (2002)  & 0.274  &  0.0470 & 1.529  & 0.0662  & 0.101 \\
$\phantom{\tablenotemark{c}}$w15.0\tablenotemark{c}         & 0.225  &  0.0495 & 1.318 & 0.0176  & 0.023 \\
w18.0         & 0.283  &  0.0430 & 1.472  & 0.0530  & 0.078 \\
w20.0         & 0.284  &  0.0393 & 1.616  & 0.0469  & 0.076 \\
n20.0         & 0.194  &  0.0580 & 1.679  & 0.0441  & 0.074
\enddata
\tablenotetext{a}{Fit parameters of separation curve
(Eq.~\ref{eq:sepcurve}) when $x$ and $y$ are measured for a
central stellar density of $5\times 10^{10}$\,g\,cm$^{-3}$.}
\tablenotetext{b}{Measured for a central stellar density of
$5\times 10^{10}$\,g\,cm$^{-3}$.}
\tablenotetext{c}{\msf \ and \scc \ measured roughly at core bounce,
because pre-collapse data are not available.}
\end{deluxetable}

\begin{figure*}[t]
\begin{center}
\includegraphics[width=\textwidth]{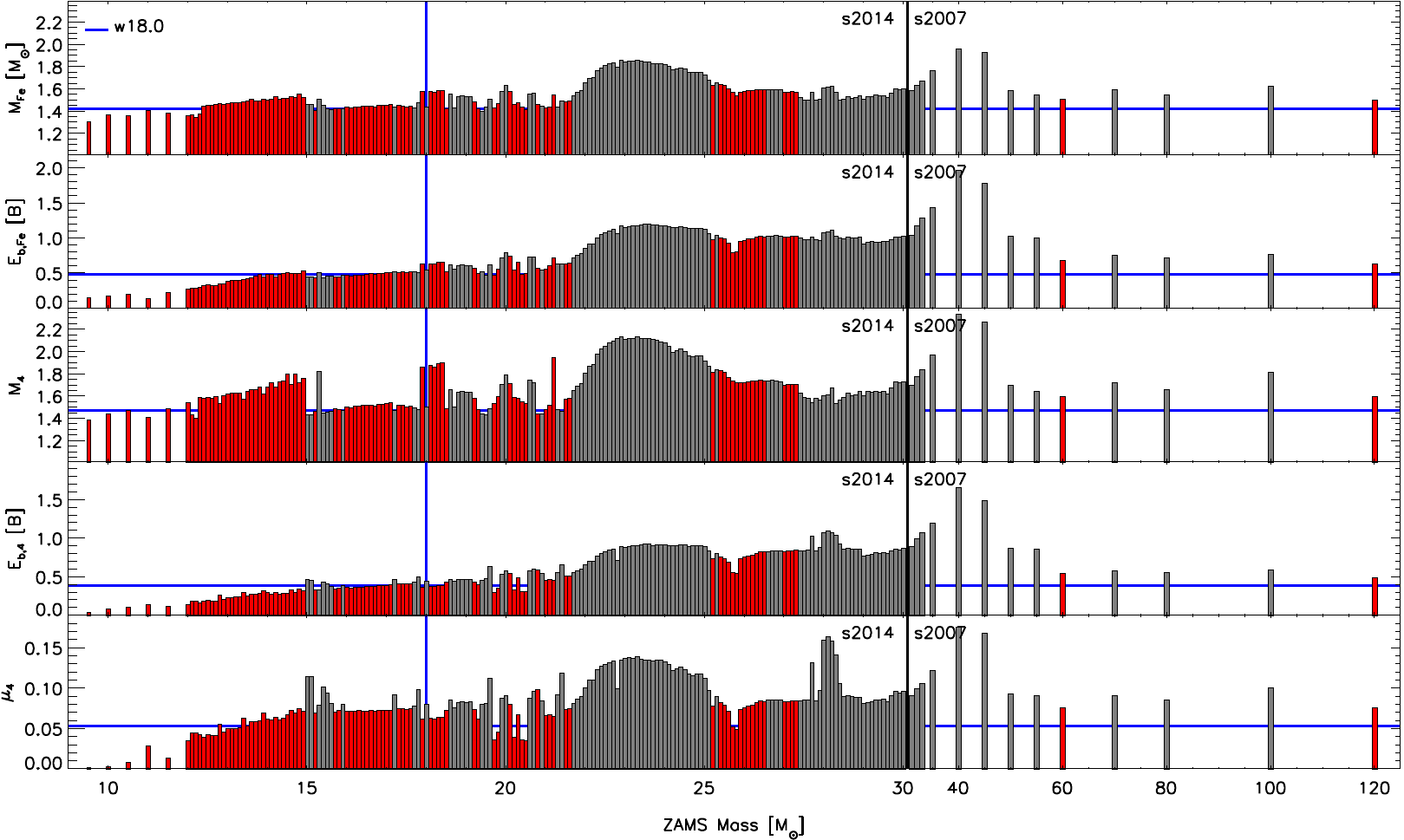}
\includegraphics[width=\textwidth]{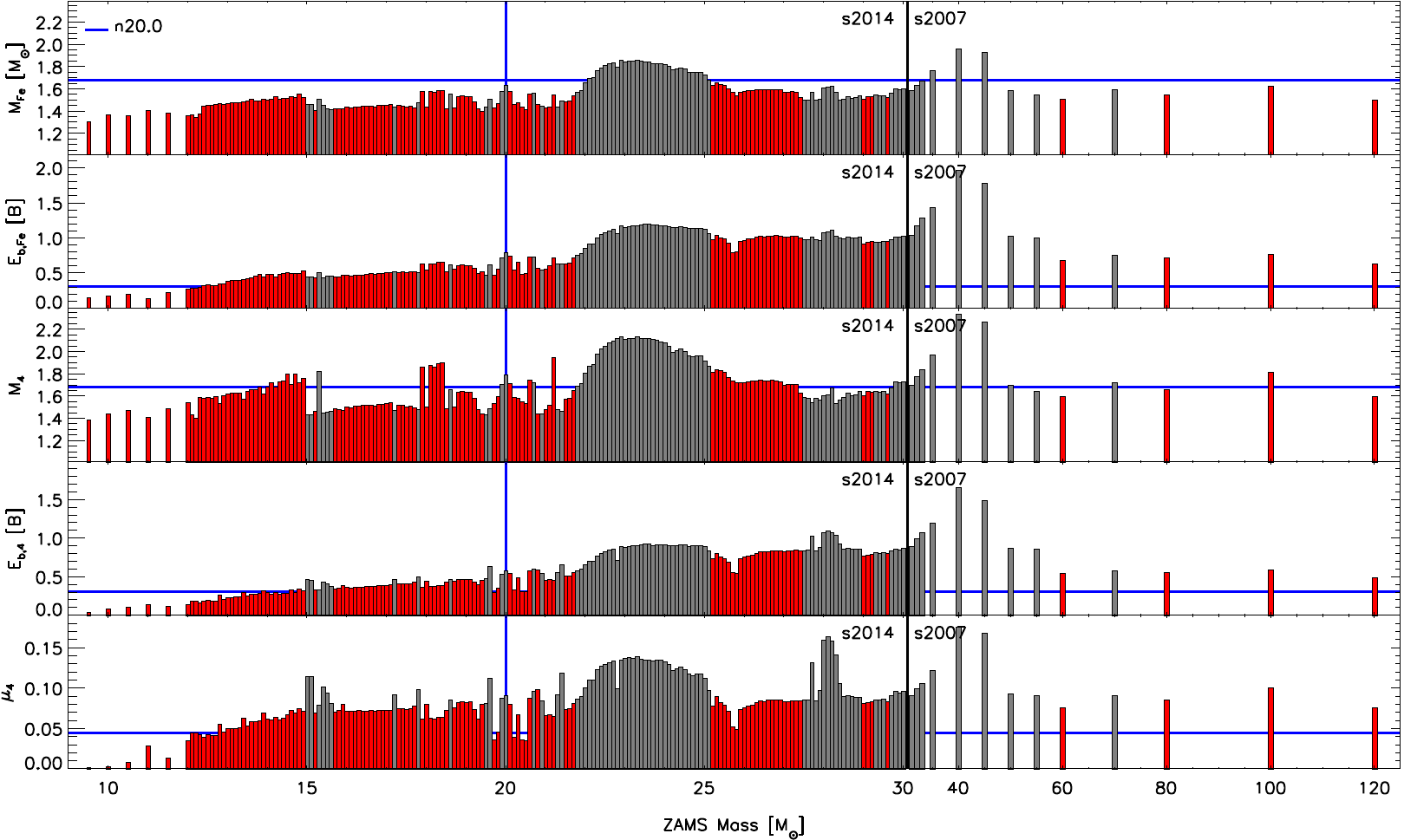}
\caption{Iron-core masses $M_\mathrm{Fe}$ ({\em top panel}),
exterior binding energies
($E_\mathrm{b,Fe}= E_\mathrm{b}(m>M_\mathrm{Fe})$; {\em second panel}),
normalized masses $M_4$ ({\em third panel}), exterior binding energies
($E_\mathrm{b,4}= E_\mathrm{b}(m/\msun>M_4)$; {\em fourth panel}), and 
$\mu_4$ ({\em fifth panel}), for models of the s2014 series, the 
supplementary low-mass progenitors with
$M_\mathrm{ZAMS}<15\,M_\odot$, and models with
$M_\mathrm{ZAMS}>30\,M_\odot$ from the s2007 series. A black
vertical line marks the boundary between the two progenitor sets.
Red bars indicate exploding cases and gray bars non-exploding ones.
All quantities are measured when the central density of the
collapsing stellar iron core is $5\times 10^{10}$\,g\,cm$^{-3}$.
The {\em upper five panels} correspond to the w18.0 calibration,
the {\em lower five panels} to the n20.0 calibration.
Mass and parameter values of the
calibration models are indicated by vertical and horizontal blue
lines, respectively. In the region of
$M_\mathrm{ZAMS}\lesssim 22\,M_\odot$ non-exploding cases, with very
few exceptions, correlate with local minima of $M_4$ and pronounced local
maxima of $\mu_4$ and $E_\mathrm{b,4}$. A high value of $M_4$ combined with 
a low value of $\mu_4$ is typically supportive for an explosion because a high
accretion luminosity (due to a high accretor mass $M_\mathrm{ns}\approx M_4$)
comes together with a low mass accretion rate (and thus low ram pressure and
low binding energy) exterior to the $s=4$ interface. The iron-core masses and
their exterior binding energies show a similar tendency, but significantly
less pronounced.
}
\label{fig:m4eb4}
\end{center}
\end{figure*}

The stellar models of all progenitor sets populate a
narrow strip in the $x$-$y$ plane of
Fig.~\ref{fig:criterion}, left panels. BH formation cases
are located in the upper left part of the $x$-$y$ plane.
The inclination of the separation line implies that the explosion
limit in terms of $\mu_4$ depends on the value of the $M_4\mu_4$
and therefore a single
parameter would fail to predict the right behavior in a large
number of cases. Denser cores outside of $M_4$ with high
mass-accretion rates (larger $\mu_4$) prevent explosions above
some limiting value. This limit grows for more massive
cores and thus higher $M_4$ because larger mass-accretion
rates do not only hamper shock expansion by higher ram pressure
but larger core masses and bigger accretion rates also correlate
with an increase of the neutrino luminosity of the PNS as expressed
by our parameter $M_4\mu_4$. The evolution tracks of successful
explosion cases in the left panel of Fig.~\ref{fig:cartoon} indicate
that for higher $L_\nu\propto M_\mathrm{ns}\dot M$ the explosion
threshold, $L_{\nu,\mathrm{crit}}(\dot M)$, can be reached for
larger values of $\dot M$.

Explosions are supported by the combination of a massive
PNS, which is associated with a high neutrino luminosity from
the cooling of the accretion mantle, on the one hand, and a rapid
decline of the accretion rate, which leads to decreasing ram pressure,
on the other hand. A high value of $M_4$ combined with a low value
of $\mu_4$ is therefore favorable for an explosion because a high
accretion luminosity (due to a high accretor mass $M_\mathrm{ns}\approx M_4$)
comes together with a low mass accretion rate (and thus low ram pressure and
low binding energy) exterior to the $s=4$ interface 
(cf.\ Fig~\ref{fig:m4eb4}). 
Such conditions are met, and explosions occur readily, when the entropy
step at the $s=4$ location is big, because a high entropy value
outside of $M_4$ correlates with low densities and a low
accretion rate. $M_4$ is usually the base of the oxygen shell
and a place where the entropy changes discontinuously causing
(or resulting from) a sudden decrease in density due to burning
there. This translates into an abrupt decrease in $\dot M$ when
the mass $M_4$ accretes. Figure~14 of \cite{Sukhbold2014} shows
a strong correlation between compactness $\xi_{2.5}$ and
location of the oxygen shell. The decrease of the mass accretion
rate is abrupt only if the entropy change is steep with mass,
for which $\mu_4$ at $M_4$ is a relevant measure.

Progenitors with $M_\mathrm{ZAMS}\lesssim 22\,M_\odot$
that are harder to explode
often have relatively small values of $M_4$ and
an entropy ledge above $s=4$ on a lower level than the entropy
reached in more easily exploding stars. The lower neutrino luminosity
associated with the smaller accretor mass in combination with
the higher ram pressure can prohibit shock expansion in many of these
cases. Corresponding to the relatively small values of $M_4$ and
relatively higher densities outside of this mass, these cases
stick out from their neighboring stars
with respect to the binding energy
of overlying material, namely, non-exploding models in almost all
cases are characterized by {\em local} maxima of
$E_\mathrm{b,4}= E_\mathrm{b}(m/\msun>M_4)$ (see Fig.~\ref{fig:m4eb4}).

\begin{figure}[t]
\begin{center}
\includegraphics[width=\columnwidth]{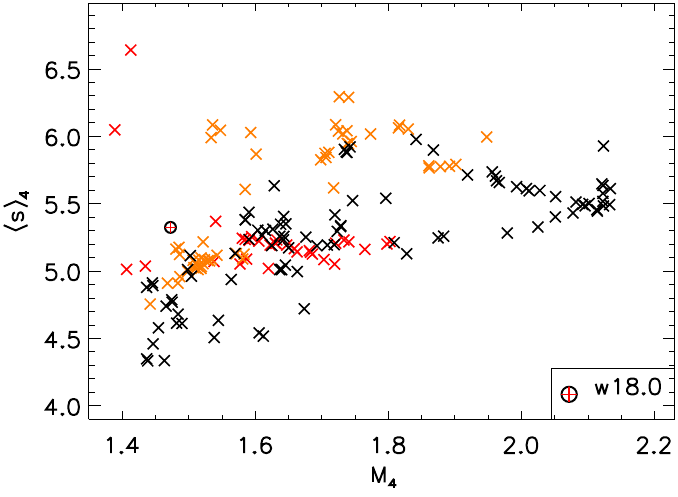}
\caption{
Two-parameter plane spanned by $M_4=M(s=4)$ and mean
entropy of overlying matter, $\langle s\rangle_4$,
averaged over a mass intervall of 0.5\,$M_\odot$.
The locations of
progenitors from the s2014 series and the supplementary
low-mass set with $M_\mathrm{ZAMS}<15\,M_\odot$ are
marked by crosses. Black crosses indicate BH formation cases
and tend to concentrate towards the lower and 
right half of the panel,
i.e., towards low values of $\langle s\rangle_4$ for given $M_4$.
Orange crosses mean successful SN explosions of the s2014
models with the w18.0 calibration, and red crosses are
explosions for the $M_\mathrm{ZAMS}<15\,M_\odot$ progenitors.
While high entropies outside of the $s=4$ location signal a
tendency of success for the stars of the s2014 set
(although the separation from BH-formation cases is not
sharp), the successfully exploding $M_\mathrm{ZAMS}<15\,M_\odot$
models mix completely with BH forming events.
}
\label{fig:m4s4}
\end{center}
\end{figure}

In view of this insight it is not astonishing that exploding
and non-exploding progenitors can be seen to start separating
from each other in the two-parameter space spanned by $M_4$
and the average entropy value $\langle s\rangle_4$ just outside
of $M_4$ (Fig.~\ref{fig:m4s4}).
Averaging $s$ over the mass interval $[M_4,M_4+0.5]$
turns out to yield the best results. Exploding models cluster
towards the side of high $\langle s\rangle_4$ and low $M_4$,
while failures are found preferentially for low values of
$\langle s\rangle_4$. The threshold for success tends to
grow with $M_4$. However, there is still a broad band
where both types of outcomes overlap. The disentanglement
of SNe and BH-formation events is clearly better achieved by
the parameter set of $M_4\mu_4$ and $\mu_4$, which, in addition,
applies correctly not only for stars with
$M_\mathrm{ZAMS}\ge 15\,M_\odot$ but also for progenitors
with lower masses.

\subsection{Stellar outliers}

Out of 621 simulated stellar models for the s19.8, w15.0, w18.0,
w20.0, and n20.0 calibrations only 9, 14, 16, 11, and 9 models,
respectively, do not follow the behavior predicted by their
locations on the one or the other side of the separation line
in the $M_4\mu_4$-$\mu_4$ plane
(see the zooms in the right column of Fig.~\ref{fig:criterion}).
But most of these cases lie very close to the boundary curve
and their explosion or non-explosion can be affected
by fine details and will certainly depend on multi-dimensional
effects. A small sample of outliers is farther away from the
boundary line. The w20.0
calibration is the weakest driver of neutrino-powered
explosions in our set and tends to yield the largest number
of such more extreme outliers.

These cases possess unusual structural features that influence
their readiness to explode.
On the non-exploding side of the separation line,
model s20.8 of the s2014 series with
$(M_4\mu_4,\mu_4)\approx(0.142,0.0981)$ is one example of a
progenitor that blows up with all calibrations except w20.0,
although it is predicted to fail (see Fig.~\ref{fig:criterion}).
In contrast, its mass-neighbor s20.9 with
$(M_4\mu_4,\mu_4)\approx(0.123,0.085)$ as well as its close
neighbor in the $M_4\mu_4$-$\mu_4$ space, s15.8 of the s2002
series with $(M_4\mu_4,\mu_4)\approx(0.140,0.096)$, both form
BHs as expected. The structure of these pre-supernova models
in the $s=4$ region is very similar with
$M_4=$\,1.45,\,1.45,\,1.46 for s20.8, s20.9, s15.8, respectively.
Although s20.8 reaches a lower entropy level outside of $s=4$
than the other two cases and therefore is also predicted to
fail, its explosion becomes possible when the next entropy
step at an enclosed mass of 1.77\,$M_\odot$ reaches the shock.
This step is slightly farther out (at 1.78\,$M_\odot$) in the
s20.9 case and comes much later (at $\sim$1.9\,$M_\odot$) in
the s15.8 model. Both the earlier entropy jump and the lower
preceding entropy level enable the
explosion of s20.8, because the associated higher density
maintains a higher mass-accretion rate and therefore higher
neutrino luminosity until the entropy jump at 1.77\,$M_\odot$
falls into the
shock. The abnormal structure of the progenitor therefore
prevents that the explosion behavior is correctly captured
by our two-parameter criterion for the explodability.

On the exploding side, model s15.3 of the s2014 series with
$(M_4\mu_4,\mu_4)\approx(0.146,0.0797)$ is expected to blow
up according to the two-parameter criterion, but does not do so
for all calibrations (Fig.~\ref{fig:criterion}). Similarly,
s15.0 of the s2007 series with $(M_4\mu_4,\mu_4)\approx(0.137,0.0749)$
fails with the w15.0 and w20.0 calibrations although success is
predicted. We compare their structure with the nearby successful
cases of s25.4 (s2014 series, $(M_4\mu_4,\mu_4)\approx(0.150,0.0820)$),
and s25.2, s25.5 (both from the s2014 series), and
s25.8 (s2002 series), all of which group around
$(M_4\mu_4,\mu_4)\approx(0.143,0.0783)$. The successfully exploding
models all have similar entropy and density structures, namely
fairly low entropies ($s\lesssim 3$) and therefore high densities
up to 1.81--1.82\,$M_\odot$, where the
entropy jumps to $s\gtrsim 6$. The high mass-accretion rate leads
to an early arrival of the $s=4$ interface at the shock
($\sim$300\,ms after bounce) and high accretion luminosity.
Together with the strong decline of the accretion rate afterwards
this fosters the explosion. In contrast, the
two models that blow up less easily have higher
entropies and lower densities so that the $s=4$ mass shells
(at $\sim$1.8\,$M_\odot$ in s15.0 and at
$\sim$1.82\,$M_\odot$ in the s15.3) arrive at
the shock much later (at $\sim$680\,ms and $\sim$830\,ms post
bounce, respectively), at which time accretion
contributes less neutrino luminosity. Moreover, both models
have a pronounced entropy ledge with a width of $\sim$0.05\,$M_\odot$
(s15.0) and $\sim$0.08\,$M_\odot$ (s15.3) before the entropy
rises above $s\sim 5$. This ledge is much narrower than in the
majority of non-exploding models, where it stretches across typically
0.3\,$M_\odot$ or more. The continued, relatively high accretion
rate prohibits shock expansion and explosion. This is obvious
from the fact that model s15.0 with the less extended entropy
ledge exhibits a stronger tendency to explode and for some
calibrations indeed does, whereas s15.3 with the wider ledge fails
for all calibrations. Our diagnostic parameter $\mu_4$ to
measure the mass derivative in an interval of
$\Delta m=0.3\,M_\odot$, however, is dominated by the
high-entropy level (low-density region) above the ledge and
therefore underestimates the mass-accretion rate in the
ledge domain, which is relevant for describing the explosion
conditions. Again the abnormal structure of the s15.0 and
s15.3 progenitors prevents our two-parameter classification
from correctly describing the explosion behavior of these models.

\begin{figure}[t]
\begin{center}
\includegraphics[width=\columnwidth]{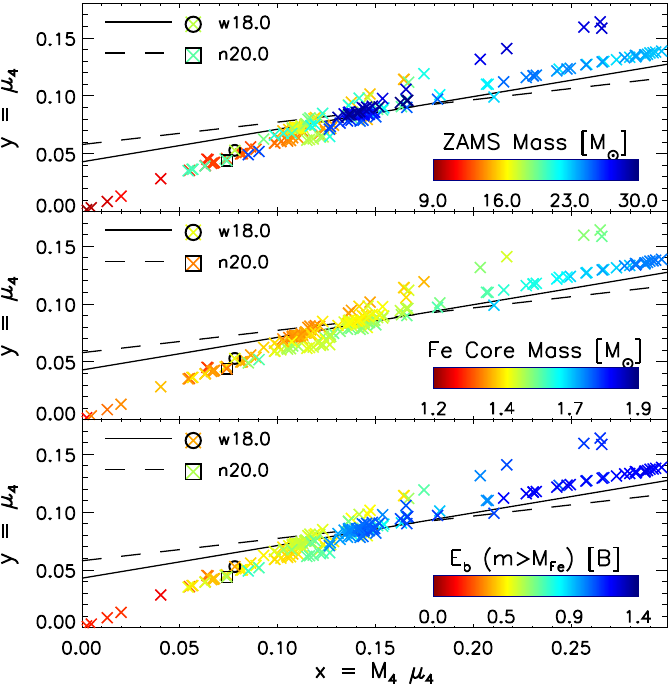}
\caption{
ZAMS masses, iron-core masses, $M_\mathrm{Fe}$, and binding energies
outside of the iron core, $E_\mathrm{b}(m>M_\mathrm{Fe})$,
({\em from top to bottom}) of the s2014 series and the supplementary
low-mass progenitors with $M_\mathrm{ZAMS}<15\,M_\odot$ in the
$x$-$y$ parameter plane. The solid
and dashed lines mark the separation curves $y_\mathrm{sep}(x)$ for
the w18.0 and n20.0 calibration models (different symbols as given in the
legend), respectively.
}
\label{fig:progproperties}
\end{center}
\end{figure}

\begin{figure*}[!]
\begin{center}
\includegraphics[width=.495\textwidth]{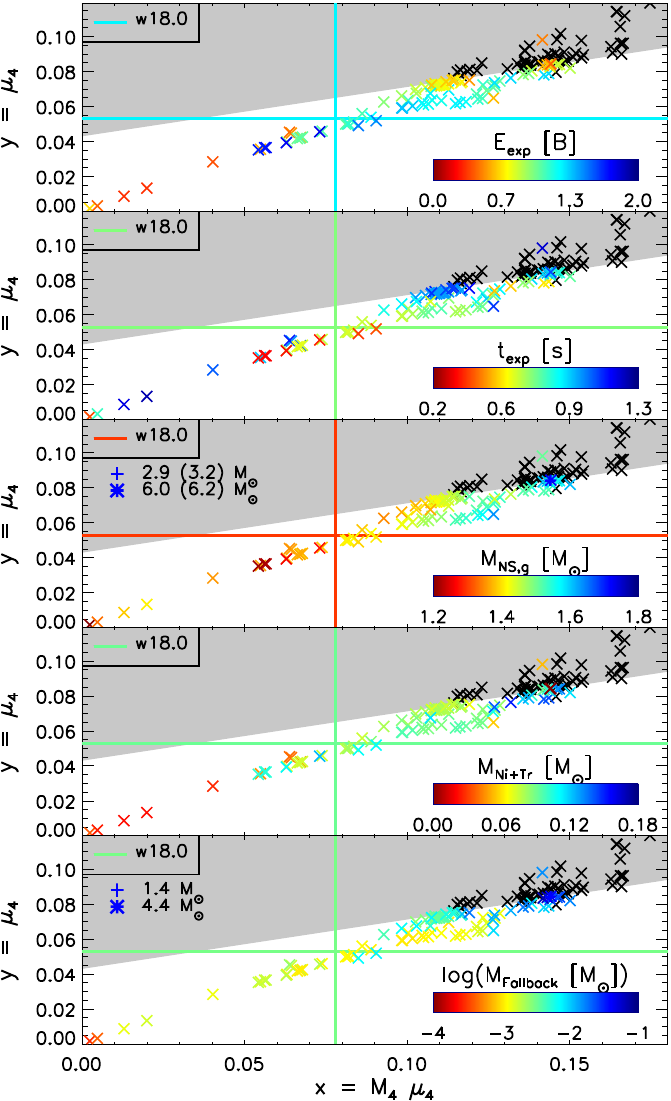}\hskip4.0pt
\includegraphics[width=.495\textwidth]{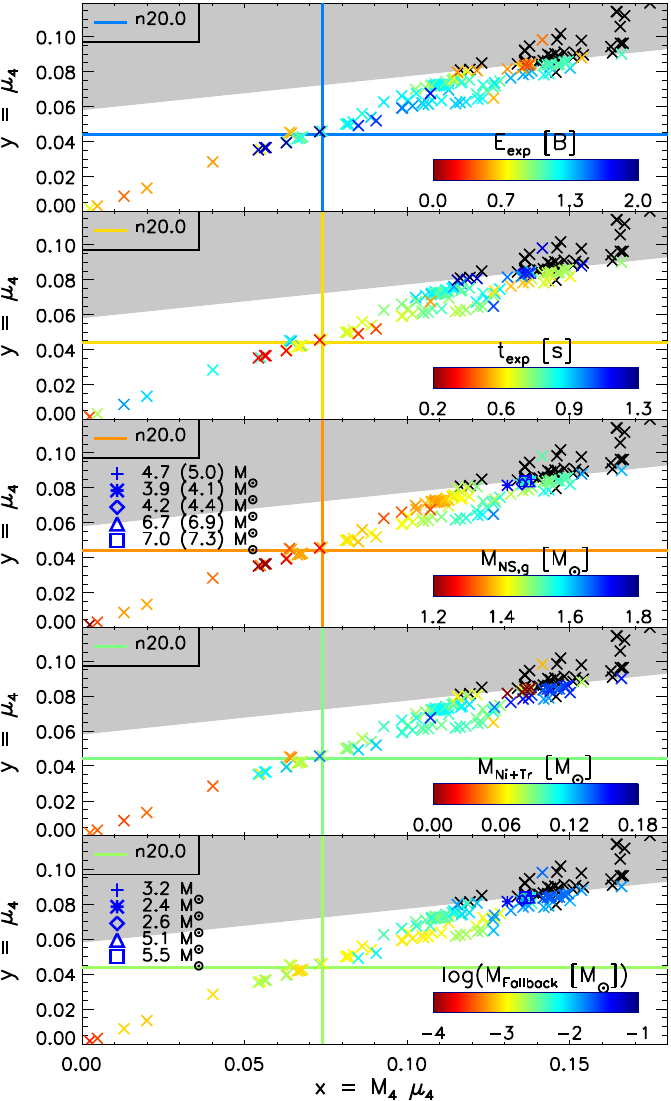}
\caption{
Explosion energies (1\,B\,=\,1\,bethe\,=\,$10^{51}$\,erg),
post-bounce explosion times, gravitational neutron-star
masses ($M_\mathrm{ns,g}=M_\mathrm{ns,b}-E_{\nu,\mathrm{tot}}/c^2$),
ejected iron-group material (i.e., $^{56}$Ni plus tracer masses), 
and fallback masses
({\em from top to bottom}) of the s2014 series and the supplementary
low-mass progenitors with $M_\mathrm{ZAMS}<15\,M_\odot$ for calibration models
w18.0 ({\em left}) and n20.0 ({\em right}) in the $x$-$y$ parameter plane.
Black crosses correspond to BH formation cases, colored crosses to successful
explosions. In the middle and bottom panels, the blue (partly overlapping)
symbols correspond to fallback SNe with estimated BH masses
(baryonic masses in parentheses) and fallback masses as listed in the
legends. The horizonal and vertical lines mark the locations of the 
calibration models with the colors corresponding to the values of the
displayed quantities.
}
\label{fig:properties}
\end{center}
\end{figure*}

\subsection{Systematics of progenitor and explosion properties
in the two-parameter plane}
\label{sec:systematics}

In Fig.~\ref{fig:progproperties} colored symbols show the
positions of the progenitors of the
s2014 series and those of the supplementary low-mass models
with $M_\mathrm{ZAMS}<15\,M_\odot$ in the $x$-$y$-plane relative
to the separation lines $y_\mathrm{sep}(x)$
of exploding and non-exploding cases.
In the upper panel the color coding corresponds to $M_\mathrm{ZAMS}$,
in the middle panel to the iron-core mass, $M_\mathrm{Fe}$,
and in the bottom panel
to the binding energy of matter outside of the iron core.
$M_\mathrm{Fe}$ is taken to be the value provided by the stellar
progenitor model at the start of the collapse simulation in
order to avoid misestimation associated with our simplified
nuclear burning network and with inaccuracies from the initial
mapping of the progenitor data. Since we use the pressure
profile of the progenitor model instead of the temperature
profile, slight differences of the derived temperatures can
affect the temperature-sensitive shell burning and thus the
growth of the iron-core mass.

While low-mass progenitors with small iron cores and low binding
energies populate the region towards
the lower left corner with significant distance to the
separation curve, stars above 20\,$M_\odot$ with bigger
iron cores and high binding energies can be mostly found well
above the separation curve. However, there are quite a
number of intermediate-mass progenitors above the line and
higher-mass cases below. In particular, a lot of stars with
masses between $\sim$25\,$M_\odot$ and 30\,$M_\odot$ cluster
around $y_\mathrm{sep}(x)$ in the $x\sim\,$0.13--0.15 region.
These stars are characterized by
$M_\mathrm{Fe}\sim\,$1.4--1.5\,$M_\odot$ and high exterior
binding energies. Some of them explode but most fail
(cf.\ Fig.~\ref{fig:m4eb4}). The ones that group on the
unsuccessful side are mostly cases with smaller iron cores, whose
neutrino luminosity is insufficient to create enough power of
neutrino heating to overcome the ram pressure of the massive
infall.

Figure~\ref{fig:properties} displays the BH-formation cases
of the s2014 series without associated SNe by black crosses
in the $x$-$y$-plane.
Successful SN explosions of this series plus additional
$M_\mathrm{ZAMS}<15\,M_\odot$ progenitors are shown by
color-coded symbols, which represent, from top to bottom,
the final explosion
energy ($E_\mathrm{exp}$, with the binding energy of the
whole progenitor taken into account), the explosion time
($t_\mathrm{exp}$, measured by the time the outgoing shock
reaches 500\,km), the gravitational mass of the remnant (with
fallback taken into account),
the ejected mass of $^{56}$Ni plus tracer element (see
Sect.~\ref{sec:numerics}; fallback also taken into account),
and the fallback mass. The left plot shows the results
of our w18.0 calibration, the right plot for the n20.0 calibration.
The gravitational mass of the NS remnant, $M_\mathrm{ns,g}$, is
estimated from the baryonic mass, $M_\mathrm{ns,b}$, by subtracting
the rest-mass equivalent of the total neutrino energy carried away
in our simulations:
\begin{equation}
M_\mathrm{ns,g}=M_\mathrm{ns,b}-\frac{1}{c^2}E_{\nu,\mathrm{tot}}\,.
\label{eq:gravmass}
\end{equation}
Our estimates of NS binding energies,
$E_\mathrm{ns,b}=E_{\nu,\mathrm{tot}}$, are roughly compatible
with \cite{Lattimer1989} fit of
$E_\mathrm{ns,b}^\mathrm{LY}=1.5\times 10^{53}\,(M_\mathrm{ns,g}/M_\odot)^2
\,\mathrm{erg}=0.084\,M_\odot c^2\,(M_\mathrm{ns,g}/M_\odot)^2$.
Blue symbols in the middle and bottom panels mark fallback BH
formation cases, for which gravitational and baryonic masses
(the latter in parentheses) are listed in the panels.
We ignore neutrino-energy losses during fallback accretion for both
NS and BH-forming remnants.

Progenitors in Fig.~\ref{fig:properties}
that lie very close to the separation line tend to
produce weaker explosions that set in later than those of
progenitors with a somewhat greater distance from the line.
Moreover, there is a tendency of more massive NSs to be produced
higher up along the $x$-$y$-band where the progenitors cluster,
i.e., bigger NS masses are made at higher values of $x=M_4\mu_4$.
Also the largest ejecta masses of $^{56}$Ni and $^{56}$Ni plus
tracer are found towards the right side of the displayed
progenitor band just below the boundary of the BH formation region.

Fallback masses tend to decrease towards the lower left corner of
the $x$-$y$-plane in Fig.~\ref{fig:properties},
far away from the separation curve, where
predominantly low-mass progenitors are located, besides five
progenitors around 20\,$M_\odot$, which lie in this region because
they have extremely low values of $E_\mathrm{b}(m>M_4)$, see
Fig.~\ref{fig:m4eb4}, and exceptionally small values of $\mu_4$
(Fig.~\ref{fig:progproperties}), and develop fast and strong explosions
with small NS masses, large masses of ejected $^{56}$Ni plus tracer,
and very little fallback.
Closer to the separation line the fallback masses are higher,
but for successfully exploding models they exceed $\sim$0.05\,$M_\odot$
only in a few special cases of fallback SNe 
(cf.\ Fig.~\ref{fig:explosionprops}), 
where the fallback mass can amount up to several solar masses.
We point out that the fallback masses in particular of stars
below $\sim$20\,$M_\odot$  were massively overestimated by
\cite{Ugliano2012}. The reason was an erroneous interpretation
of the outward reflection of reverse-shock accelerated matter as a
numerical artifact connected to the use of the condition at the
inner grid boundary. The
reverse shock, which forms when the SN shock passes the He/H interface,
travels backward through the ejecta and decelerates the outward moving
matter to initially negative velocities. This inward flow of
stellar material, however, is slowed down and reflected back outward by
the large negative pressure gradient that builds up in the reverse-shock
heated inner region. With this outward reflection, which is a true
physical phenomenon and not a boundary artifact, the matter that
ultimately can be accreted by the NS is diminished to typically between
some $10^{-4}$\,$M_\odot$ and some $10^{-2}$\,$M_\odot$ of early fallback
(see Fig.~\ref{fig:explosionprops}).

In a handful of high-mass s2014 progenitors (s27.2 and s27.3
for the w18.0 calibation and s27.4, s29.0, s29.1, s29.2, and s29.6
for the n20.0 set) the SN explosion is unable to unbind a large
fraction of the star so that fallback of more than a solar mass
of stellar matter is likely to push the NS beyond the BH formation
limit. Such fallback SN cases cluster in the vicinity of
$x\sim\,$0.13--0.14 and $y\sim\,$0.080--0.081 on the explosion
side of the separation line. In the ZAMS-mass sequence they lie at
interfaces between mass intervals of successfully exploding and
non-exploding models, or they
appear isolated in BH-formation regions of the
ZAMS-mass space (see Figs.~\ref{fig:explosions} and \ref{fig:m4eb4}).
Their fallback masses
and estimated BH masses are listed in the corresponding panels
of Fig.~\ref{fig:properties}. Naturally, they stick out also by their
extremely low ejecta masses of $^{56}$Ni and tracer elements, late
explosion times (around one second post bounce or later), and
relatively low explosion energies ($\sim$0.3--0.5\,B).

\newpage
\subsection{Influence of the neutron star radius}
\label{sec:nsradius}

Basically, the accretion luminosity, which is given by
$L_\nu^\mathrm{acc}\propto GM_\mathrm{ns}\dot M/R_\mathrm{ns}$,
does not only depend on the PNS mass, $M_\mathrm{ns}$, and the
mass-accretion rate, $\dot M$, but also on the PNS radius, 
$R_\mathrm{ns}$. One may wonder whether our two-parameter criterion
is able to capture the essential physics although we disregard
the radius dependence when using $x = M_4\mu_4$ as a proxy of
the accretion luminosity.

For this reason, we tested a redefinition of $x = M_4\mu_4$ by 
including a factor $\langle R_\mathrm{ns}\rangle_{300\,\mathrm{ms}}^{-1}$,
i.e., using $\tilde{x}\equiv M_4\mu_4 \langle
R_\mathrm{ns}\rangle_{300\,\mathrm{ms}}^{-1}$ instead\footnote{The radius
$R_\mathrm{ns}$ of the proto-neutron star is defined by the radial
position where the density is $10^{11}$\,g\,cm$^{-3}$. As in panels
d--f of Fig.~\ref{fig:correlations}, the time-averaging for
$\langle R_\mathrm{ns}\rangle_{300\,\mathrm{ms}}^{-1}$ is performed 
from the passage of the $s=4$ interface through the shock until 
the onset of the explosion for successful models and from the 
infall of the $s=4$ interface until 300\,ms later for non-exploding
models. We also employ a normalization factor of 70\,km to recover 
(roughly) the same range of values for $\tilde{x}$ as in the case
of $x$.}. 
Doing so, we found essentially no relevant effects on the location
of the boundary curve. In fact, the separation of exploding and 
non-exploding models in the $\tilde{x}$-$y$-plane is even slightly
improved compared to the $x$-$y$-plane, because 
$\langle R_\mathrm{ns}\rangle_{300\,\mathrm{ms}}^{-1}$ for most
non-exploding models is larger than for the far majority of
exploding ones. As a consequence, the non-exploding cases tend to be 
shifted to the left away from the boundary line, whereas most of the
exploding cases are shifted to the right, also increasing their
distances to the boundary line. This trend leads to a marginally
clearer disentanglement of both model groups near the border between
explosion and non-explosion regions. A subset of the (anyway 
few and marginal)
outliers can thus move to the correct side, while very few cases
can become new, marginal outliers. It might therefore
even be possible to improve the success rate for the classification
of explodability by a corresponding (minor) relocation of the boundary 
curve. The improvement, however, is not significant enough to 
justify the introduction of an additional parameter 
into our two-parameter criterion in the form of 
$\langle R_\mathrm{ns}\rangle_{300\,\mathrm{ms}}$, which has the
disadvantage of not being based on progenitor properties and whose
exact, case-dependent value cannot be predicted by simple arguments.

Because the line separating exploding and non-exploding models did
not change in our test, the criterion advocated in this paper captures
the basic physics, within the limitations of the modeling.
While we report here this marginal sensitivity of our two-parameter 
criterion to the NS radius as a result of the
present study, a finally conclusive assessment of this
question would require to repeat our set of model calculations
for different prescriptions of the time-dependent contraction of
the inner boundary of our computational grid. The chosen functional
behavior of this boundary movement with time determines how the 
proto-neutron star contraction proceeds during the crucial phase
of shock revival. In order to avoid overly severe numerical 
time-step constraints, which can become a serious handicap
for our long-time simulations with explicit neutrino transport
over typically 20 seconds, we follow \cite{Ugliano2012} in using
a relatively slow contraction of the inner grid boundary. It
would be highly desirable to perform model calculations also
for faster boundary contractions, which is our plan for future
work. In view of this caveat the arguments and test results 
discussed in this section should still be taken with a grain
of salt.

On grounds of the discussion of our results in the $x$-$y$-plane
one can actually easily understand why the
definition of the separation curve of exploding and non-exploding
models in the present paper did not require us to take into
account a possible
dependence of the accretion luminosity on the NS radius.
Instead, we could safely ignore such a dependence when we
coined our ansatz that $L_\nu \propto Mm'(r)\propto M_4\mu_4$.
There are two reasons for that. On the one hand, the
separation line in the $M_4\mu_4$-$\mu_4$ plane is fairly
flat. A variation of the NS
radius corresponds to a horizontal shift of the location
of data points in the $x$-$y$ plane. However, 
with the contraction behavior of the neutron stars obtained
in our simulations, only for (relatively few) points in the
very close vicinity of the separation curve such a horizontal
shift is sufficiently big to potentially have an
influence on whether the models are classified as non-exploding
or exploding. On the other hand, the models near the separation
line typically blow up fairly late ($t_\mathrm{exp}\gtrsim 0.6$\,s),
and the NS masses lie in a rather narrow range between roughly
1.4\,$M_\odot$ and 1.6\,$M_\odot$ for the gravitational mass
(see Fig.~\ref{fig:properties}). For such conditions the variation
of the NS radius is of secondary importance (cf.\ Fig.~3 in
\citealt{Muller2014}). Low-mass progenitors with less massive
NSs, whose radius (at the same time) can be somewhat larger
(Fig.~3 in \citealt{Muller2014}), however, are located towards
the lower left corner of the $x$-$y$ plane
and therefore far away from the separation curve (compare
Figs.~\ref{fig:progproperties} and \ref{fig:properties}).
An incorrect horizontal placement
of these cases (due to the omission of the dependence of
the neutrino luminosity on the NS radius) does not have any
relevance for the classification of these models.

\subsection{Brief comparison to previous works}

Our result of a complex pattern of NS and BH forming cases
as function of progenitor mass was previously found by
\cite{Ugliano2012}, too, and was confirmed by \cite{Pejcha2015}.
Aside from differences in details depending on the use of 
different progenitor sets and different SN~1987A calibration
models, the main differences of the results presented here
compared to those of \cite{Ugliano2012} are lower explosion
energies for progenitors with $M\lesssim 13\,M_\odot$ 
(see discussion in Sect.~\ref{sec:calibration}) and lower 
fallback masses as mentioned in Sect.~\ref{sec:systematics}.
Based on simple arguments (which, however, cannot account
for the complex dynamics of fallback), \cite{Pejcha2015}
already expected (in particular for their parameterization (a)) 
that cases with significant fallback ---in the sense that
the remnant masses are significantly affected--- should be rare 
for solar-metallicity progenitors. Our results confirm this
expectation, although the ZAMS masses with significant
fallback are different and less numerous than in the work 
by \cite{Pejcha2015}. As pointed out by these authors, 
fallback has potentially important consequences for the 
remnant mass distribution, and the observed NS and BH
masses seem to favor little fallback for the majority
of SNe.

As discussed in detail by \cite{Ugliano2012}, our explosion
models (as well as parameterization (a) of \citealt{Pejcha2015})
predicts many more BH formation cases and more mass intervals 
of non-exploding stars than \cite{OConnor2011}, who made
the assumption that stars with compactness 
$\xi_{2.5}> 0.45$ do not explode. Roughly consistent with our 
results and those of \cite{Ugliano2012}, \cite{Horiuchi2014}
concluded on the basis of observational arguments (comparisons
of the SN rate with the star formation rate; the red supergiant
problem as a lack of Type-IIP SNe from progenitors above a
mass of $\sim$16\,$M_\odot$) that stars with an ``average'' 
compactness of $\xi_{2.5}\sim 0.2$ should fail to produce 
canonical SNe (cf.\ Fig.~15 in \citealt{Sukhbold2015}).

A correlation of explosion energy and $^{56}$Ni mass as
found by \cite{Pejcha2015} and \cite{Nakamura2015} 
(both, however, without rigorous determination of 
observable explosion energies at infinity) and as suggested
by observations (see \citealt{Pejcha2015} for details)
is also predicted by our present results (but not by 
\citealt{Ugliano2012} with their erroneous estimate of the
fallback masses and the overestimated explosion energies of 
low-mass progenitors, see Sect.~\ref{sec:calibration}).
Our models yield low explosion
energies and low nickel production towards the low-mass
side of the SN progenitors (see also \citealt{Sukhbold2015}).
In contrast, the modeling approach by \cite{Pejcha2015}
predicts a tendency of lower explosion energies and lower
$^{56}$Ni masses towards the high-mass side of the 
progenitor distribution (because of the larger binding
energies of more massive stars), although there is a large
mass-to-mass scatter in all results. This difference
could be interesting for observational diagnostics.
The modeling approach by \cite{Pejcha2015} seems to
yield neutrino-driven winds that are considerably stronger,
especially in cases of low-mass SN progenitors,
than those obtained in the simulations of \cite{Ugliano2012}
and, in particular, than those found in current, fully 
self-consistent SN and PNS cooling models, whose behavior we
attempt to reproduce better by the Crab-motivated recalibration
of the low-mass explosions used in the present work.
Moreover, for the $^{56}$Ni-$E_\mathrm{exp}$ correlation
reported by \cite{Pejcha2015},
a mass interval between $\sim$14\,$M_\odot$ and
$\sim$15\,$M_\odot$ with very low explosion energies and
very low Ni production, which does not exist in our models,
also plays an important role.
\cite{Nakamura2015} found a positive correlation of the
explosion energy and the $^{56}$Ni mass with compactness
$\xi_{2.5}$. We can confirm this result for the nickel
production but not for the explosion energy (cf.\ 
\citealt{Sukhbold2015}, Fig.~15 there). A possible reason for
this discrepancy could be the consideration of ``diagnostic''
energies by \cite{Nakamura2015} at model dependent final
times of their simulations instead of asymptotic explosion
energies at infinity, whose determination requires seconds
of calculation and the inclusion of the binding energy of
the outer progenitor shells (see Fig.~\ref{fig:explosionprops}
and \citealt{Sukhbold2015}, Fig.~6, for the evolution of the
energies in some exemplary simulations).

Since a comprehensive discussion of explosion energies,
nickel production, and remnant
masses is not in the focus of the present work, we refer the
reader interested in these aspects to the follow-up paper by
\cite{Sukhbold2015}. For the same reason we also refrain here
from more extended comparisons to the progenitor-dependent
explosion and remnant predictions of other studies,
in particular those of \cite{OConnor2011}, \cite{Nakamura2015},
and \cite{Pejcha2015}.
A detailed assessment of the different modeling methodologies 
and their underlying assumptions
would be needed to understand and judge the reasons for
differences of the results and to value their meaning in the
context of supernova predictions based on the neutrino-driven 
mechanism. Such a goal reaches far beyond the scope of our
paper.

\section{Conclusions}
\label{sec:conclusions}

We performed 1D simulations of SNe for a large set of
621 progenitors of different masses and metallicities, including
the solar-metallicity s2002 series \citep{Woosley2002}
previously investigated by \cite{Ugliano2012} and the
new s2014 models of \cite{Sukhbold2014} with their fine
gridding of 0.1\,$M_\odot$ in the ZAMS mass.

In order to obtain SN explosions in spherical symmetry, we adopt
the methodology of \cite{Ugliano2012}, using 1D hydrodynamics
and approximate neutrino transport and a PNS-core neutrino source,
but with improvements
in a number of modeling aspects (e.g., a nuclear high-density EoS
and a fully self-consistent implementation of nuclear burning
through a small network, cf.\ Sect.~\ref{sec:modeling}).
Explosions are triggered by a neutrino luminosity
that is sufficiently large to overcome the critical threshold
condition for shock runaway.
This luminosity is fed by a progenitor-dependent
accretion component during the post-bounce shock-stagnation phase
as well as a component from the high-density core of the nascent
NS. The core emission is also progenitor-dependent, because it
scales with the mass of the hot accretion mantle that assembles around
the cooler high-density core of the PNS
during the pre-explosion evolution. The conditions for
an explosion are thus tightly coupled to the progenitor structure,
which determines the post-bounce accretion history.

Our approach contains a number of free parameters, whose values are
calibrated by reproducing observational properties (explosion
energy, $^{56}$Ni mass, total neutrino energy and signal duration)
of SN~1987A with suitable progenitor models of this SN. We consider
five different such progenitors for our study, namely besides the
s19.8 star of the s2002 model series of \cite{Woosley2002},
which was used by \cite{Ugliano2012} as calibration model,
also 15, 18, and 20\,$M_\odot$ of Woosley and collaborators as
well as a 20\,$M_\odot$ model from \cite{Shigeyama1990}
(see Sect.~\ref{sec:progenitors}).

Because 1D simulations cannot properly reproduce the period of
continued accretion and simultaneous outflow that characterizes
the early expansion of the revived shock in multi-dimensional
simulations and delivers the explosion energy,
our 1D models exhibit an extended episode of accretion
that is followed by a strong early wind phase. The overestimated
mass loss during the latter phase compensates for the enhanced
preceding accretion, and the associated recombination energy
yields the dominant contribution to the power supply of the
beginning explosion. A detailed discussion of our methodology
can be found in Sect.~\ref{sec:methodology}.

Overall, our results confirm the ZAMS-mass dependent explosion
behavior that was found by \cite{Ugliano2012} for the
s2002 model series. For the same explosion calibration this
progenitor set and the newer s2014 models have basic features
in common. Moreover, for all calibration cases we observe
similar irregular patterns of successful explosions alternating
with BH formation events above $\sim$15\,$M_\odot$.
The largest fraction of BH
formation cases is obtained with the w20.0 calibration model,
a 20\,$M_\odot$ SN~1987A progenitor of \cite{Woosley1997},
which explodes relatively easily and reproduces the SN~1987A
$^{56}$Ni yield with a fairly low ratio of explosion energy to
ejecta mass of $E_\mathrm{exp}/M_\mathrm{ej}\sim 0.7$ only.
The core neutrino source is correspondingly weak and enables
successful SNe in a smaller subset of progenitors.
On the other hand, we obtain the closest similarity of the
explodability of the investigated progenitor sets when we use
the s19.8 (Ugliano et al.'s) calibration model and \cite{Shigeyama1990}
n20.0 SN~1987A progenitor, which
possess very similar compactness values $\xi_{2.5}$.

Non-exploding cases tend to correlate with {\em local} maxima
of the compactness $\xi_{2.5}$, of the total binding energy outside
of the iron core, $E_\mathrm{b}(m>M_\mathrm{Fe})$, and, most
significantly, {\em local} maxima of the total binding energy
$E_\mathrm{b}(m/\msun>M_4)$ outside of the mass coordinate
$M_4 = m(s=4)/\msun$, where the dimensionless entropy per nucleon
reaches a value of 4. Many (but not all) non-exploding progenitors
below $\sim$22\,$M_\odot$ also coincide with local minima of
$M_\mathrm{Fe}$ and, in particular, with minima of $M_4$.
However, there are no fix threshold values of any
of these characteristic parameters of the pre-collapse progenitor
structure that could be used to discriminate favorable from
non-favorable conditions for an explosion.

Guided by such insights we propose a two-parameter criterion
to classify the explodability of progenitor stars by the
neutrino-heating mechanism in dependence of the pre-collapse
properties of these stars. The two structural parameters that turn
out to yield the best separation of successful and unsuccessful
cases are $M_4$ and the mass derivative $\mu_4 =
\mathrm{d}m/\mathrm{d}r[M_\odot/(1000\,\mathrm{km})]^{-1}|_{s=4}
= m'(s=4)[M_\odot/(1000\,\mathrm{km})]^{-1}$ 
just outside of the $s=4$ location, which we combine to a
parameter $x\equiv M_4\mu_4$. The parameters $x$ and
$y\equiv \mu_4$ are tightly connected to the two
crucial quantities that govern the physics of the neutrino-driven
mechanism, namely the mass-accretion rate of the stalled shock,
$\dot M$, and the neutrino luminosity $L_\nu$. The former determines
the ram pressure that damps shock expansion and can be
mathematically linked to the mass derivative $m'$ (see
Eq.~\ref{eq:macc}). The latter is a crucial ingredient for the
neutrino heating that is responsible for shock revival and is
dominated by the accretion luminosity and the
PNS-mantle cooling emission during the crucial phase
of shock revival. Both of these scale with $\dot M$ and/or the
accretor mass (i.e., the PNS mass) so that
$L_\nu\propto M_\mathrm{ns}\dot M$ expresses the leading dependence.
Since
the neutrino-driven explosions in our simulations set in shortly
after the entropy interface and density jump around $s=4$ have
fallen through the shock (Fig.~\ref{fig:expl-examples}),
$M_4$ can be taken as a good proxy of $M_\mathrm{ns}$ as
the accretor mass, and $\mu_4$ can serve as a measure for
the mass accretion rate $\dot M$ of the PNS.

Higher values of $M_4$ tend to be favorable for an explosion as shown
by Fig.~\ref{fig:m4eb4}, where many non-exploding cases (dark gray)
correlate with local minima of $M_4$. The reason is that the
neutrino luminosity scales (roughly) with $x=M_4\mu_4$ (the actual
sensitivity of the neutrino-energy deposition to $M_4$ is even
steeper). Therefore higher $M_4$ imply greater neutrino
luminosities and stronger neutrino heating.
In contrast, the influence of $y=\mu_4$ is ambivalent. On the one
hand a high value of $\mu_4$ increases the neutrino luminosity,
on the other hand it also causes a large ram pressure that has
to be overcome by neutrino heating. The effect of these competing
influences is that a higher value of $M_4$ in association with
a lower value of $\mu_4$ favors explosions. Reversely, non-exploding
cases in Fig.~\ref{fig:m4eb4} are correlated with local minima of
$M_4$ and maxima of $\mu_4$. Visually, this means
that explosion cases are preferentially located toward
the lower right of the $M_4\mu_4$-$\mu_4$ parameter space
(cf.\ Figs.~\ref{fig:cartoon}, \ref{fig:criterion},
\ref{fig:properties}).

The parameters $x$ and $y$ therefore span a plane in which
successful explosions and failures with BH formation are clearly
separated. The progenitor stars populate an astonishingly narrow
band that stretches from the lower left corner with the lightest
stars to the upper right direction of the $x$-$y$-plane, where
the massive progenitors with the biggest iron cores and highest
binding energies of overlying material are located (see
Fig.~\ref{fig:progproperties}).
While SNe can be found in the region of low values of
$y$, i.e., for low mass-accretion rate,
the non-exploding cases inhabit the domain of high $y$,
but the limiting value of the mass-accretion rate that prevents
the success of the neutrino-driven mechanism grows with the
value of $x$. Both sectors in the $x$-$y$ plane are separated by
a boundary line that can be well represented by a linear function
$y_\mathrm{sep}(x)$ (Eq.~\ref{eq:sepcurve}) with increasing
slope. (The values of the dimensionless coefficients of this
linear relation are listed for all calibration models in
Table~\ref{table:separation}.) Because of the physical meaning
of the parameters $x$ and $y$, there is a close correspondence
between the separation line $y_\mathrm{sep}(x)$ and the critical
threshold luminosity $L_{\nu,\mathrm{crit}}(\dot M)$ that has to be
exceeded to trigger runaway expansion of the accretion shock by
neutrino heating. The rising slope of the separation curve in this
context means that for each value of the neutrino luminosity,
respectively $x$, there is an upper limit of the mass accretion
rate, respectively $y$, up to which neutrino-driven explosions
are possible, and that this BH-formation threshold value of $y$
grows with $x$. The parameters $x$ and $y$, computed from the
progenitor profiles before collapse, allow one to judge whether
a considered star is able to overcome the threshold neutrino
luminosity for an explosion, or, in other words, whether its mass
accretion rate stays below the critical limit above which the
onset of the explosion is prevented.

Only $\sim$1--3\% of all investigated
progenitors do not follow this discrimination scheme
in their final fate but lie on the wrong side of the separation
curve. These outliers are characterized by pathologies of their
entropy and density profiles that describe the composition-shell
structure in the Si-O-layers. Such special conditions lead to
unusual combinations
of mass accretion rate and PNS masses.
Our two-parameter criterion expressed by the separation line
$y_\mathrm{crit}(x)$ therefore enables one, with a very high
significance, to predict the explodability of
progenitor stars via the neutrino-driven mechanism by referring
to the properties of these stars as captured by the pair
of parameters $x$ and $y$.

\cite{Clausen2015} explore the interesting possibility that
the death of massive stars in NS vs.\ BH formation may 
be better captured by a probabilistic description.
The non-monotonic variations of explosion vs.\ non-explosion
with ZAMS mass or compactness might be interpreted as
stochasticity in the explosion behavior.
However, by considering the problem in a more appropriate
two-parameter space, our two-parameter criterion unmasks 
these putatively random variations as actually deterministic
phenomenon.
\cite{Clausen2015}, in contrast, suggest a variety of factors
besides ZAMS mass and metallicity, e.g.\ rotation, binarity,
the strength of magnetic fields, stochastic
differences in the pre-collapse structure or even in the
explosion mechanism, that might introduce a randomness such that
a star of given mass might not form either a NS or a BH 
but both with a certain probability. If such a diversity in
the stellar destiny depends on a causal process, for example
the presence of different amounts of spin, 
a deterministic description could still
apply but would require an extension to a parameter space of
more than the two dimensions combined by our current criterion, 
e.g., by adding extra dimension(s) that capture the role of
rotation in the explosion mechanism. If, in contrast, truly
stochastic effects like turbulent processes or chaotic 
fluctuations in the progenitor, decide about the stellar fate, 
a deterministic criterion for explodability would be ruled out
and a probabilistic description would be indispensable.

Our study has a number of more caveats that need to be
addressed. The understanding of the explosion mechanism(s) of
massive stars is still incomplete, although considerable progress
has been achieved in recent years due to the progress in 2D and
3D modeling and in particular also through improvements in the
treatment of the crucial neutrino physics and transport in
collapsing stellar cores
\citep[see, e.g.,][and references therein]{Janka2012a,
Janka2012b,Burrows2013,Mezzacappa2015,Melson2015,Melson2015b,Lentz2015}.
Without self-consistent 3D explosion
simulations for large sets of progenitor stars being possible
yet, our study refers to a 1D modeling approach, in which not
only the neutrino description is approximated in many aspects,
but also the explosions have to be triggered artificially.
We employ a boundary condition that replaces the
high-density, low-entropy core of the nascent NS as a
neutrino source, and we
describe the time-dependent behavior of this core and of
the coupling between the core and its hot accretion mantle
by a simple, analytical model. Calibrating the involved free
parameters by observational constraints from SN~1987A for the
more massive stars and by comparison to results of sophisticated
SN models for low-mass progenitors is intended to anchor our
simplified description on empirical ground. Although the elements
of this approximate approach appear qualified to capture the
essence of the neutrino-heating mechanism in dependence of
the progenitor-specific post-bounce accretion evolution
(cf.\ Sect.~\ref{sec:methodology} for details), confirmation by
fully self-consistent, multi-dimensional SN calculations is
ultimately indispensable. It is also
evident that our study, which is only concerned with neutrino-driven
explosions, cannot yield any information about the possibility and
implications of other mechanisms to blow up stars, for example
magnetorotational explosions of rapidly rotating stellar cores,
which might be a consequence of magnetar or BH formation and
could be associated with hypernovae and gamma-ray burst SNe
\citep[see, e.g.,][]{Mazzali2014} as well as ultra-luminous SNe
\citep{Woosley2010,Kasen2010,Sukhbold2015}.

Our study employs pre-collapse models that emerge from 1D
stellar evolution calculations of single, non-rotating SN
progenitors with prescribed mass-loss histories.
The results of our study naturally
depend on the post-bounce accretion properties of the
collapsing stars. The (iron and low-entropy) core masses as
well as the entropy and density jumps at the composition-shell
interfaces play an important role in setting the conditions
for the neutrino-heating mechanism, which is obvious from the
definition of our parameters $x$ and $y$. It is conceivable that
multi-dimensional hydrodynamics could lead to considerable
changes of the stellar properties as functions of the progenitor
mass \citep[e.g.,][]{Arnett2015}, and that asymmetries and
perturbations in the shell-burning layers of the pre-collapse
core might have important consequences for the development
of SN explosions by the neutrino-driven mechanism
\citep[e.g.,][]{Arnett2011,Couch2013b,Couch2015a,Muller2015,Couch2015b}.
We
are hopeful that the basic insights of our study, in particular
the existence of a two-parameter criterion for the explodability
---expressed by a SN-BH separation line $y_\mathrm{crit}(x)$
in the $x$-$y$-space and the tight connection between this curve
and the critical luminosity condition $L_{\nu,\mathrm{crit}}(\dot M)$
of the neutrino-driven mechanism--- possess more general validity.
The explosion properties of the progenitor stars as functions of
the ZAMS mass, however, do not only depend (moderately) on the
considered SN~1987A progenitor models but will probably also change
once multi-dimensional stellar evolution effects will be accounted
for in the pre-supernova conditions.

\acknowledgements
We thank A.~Wongwathanarat for numerical support and
R.~Hix and F.-K.~Thielemann for providing the NSE solver
used by K.~Kifonidis. We are also grateful to A.~Heger for
pointing out to us inconsistencies in the notation, and
to D.~Clausen, B.~M\"uller, and A.~Perego for comments on
the arXiv version.
At Garching, funding by Deutsche Forschungsgemeinschaft
through grant EXC 153 and the European Union through grant
ERC-AdG No.~341157-COCO2CASA is acknowledged.
At Santa Cruz, this work was supported by NASA (NNX14AH34G)
and the UC Office of the President (12-LR-237070).
At Darmstadt, this work was funded by the  Helmholtz-University
Young Investigator grant No.\ VH-NG-825.

\bibliographystyle{apj}

\end{document}